\documentclass[twocolumn]{aastex631}
\shorttitle{NGC~6316}
\shortauthors{Deras D.}
\graphicspath{{./}{figures/}}

\begin{document}

\title{Digging into the Galactic Bulge: stellar population and structure of the poorly studied cluster NGC~6316}

\author[0000-0002-3730-9664]{Dan Deras}
\affil{Dipartimento di Fisica e Astronomia, Università di Bologna, Via Gobetti 93/2 I-40129 Bologna, Italy}

\author[0000-0002-5038-3914]{Mario Cadelano}
\affil{Dipartimento di Fisica e Astronomia, Università di Bologna, Via Gobetti 93/2 I-40129 Bologna, Italy}
\affil{INAF-Osservatorio di Astrofisica e Scienze dello Spazio di Bologna, Via Gobetti 93/3 I-40129 Bologna, Italy}

\author[0000-0002-2165-8528]{Francesco R. Ferraro}
\affil{Dipartimento di Fisica e Astronomia, Università di Bologna, Via Gobetti 93/2 I-40129 Bologna, Italy}
\affil{INAF-Osservatorio di Astrofisica e Scienze dello Spazio di Bologna, Via Gobetti 93/3 I-40129 Bologna, Italy}

\author[0000-0001-5613-4938]{Barbara Lanzoni}
\affil{Dipartimento di Fisica e Astronomia, Università di Bologna, Via Gobetti 93/2 I-40129 Bologna, Italy}
\affil{INAF-Osservatorio di Astrofisica e Scienze dello Spazio di Bologna, Via Gobetti 93/3 I-40129 Bologna, Italy}

\author[0000-0002-7104-2107]{Cristina Pallanca}\affil{Dipartimento di Fisica e Astronomia, Università di Bologna, Via Gobetti 93/2 I-40129 Bologna, Italy}
\affil{INAF-Osservatorio di Astrofisica e Scienze dello Spazio di Bologna, Via Gobetti 93/3 I-40129 Bologna, Italy}



\begin{abstract}
High-resolution Hubble Space Telescope optical observations have been used to analyze the stellar population and the structure of the poorly investigated bulge globular cluster NGC~6316. We constructed the first high-resolution reddening map in the cluster direction, which allowed us to correct the evolutionary sequences in the color magnitude diagram (CMD) for the effects of differential reddening. A comparison between the CMDs of NGC 6316 and 47 Tucanae revealed strikingly similar stellar populations, with the two systems basically sharing the same turn-off, sub-giant branch, and horizontal branch morphologies, indicating comparable ages. The red giant branch in NGC 6316 appears slightly bluer than in 47 Tucanae, suggesting a lower metal content. This has been confirmed by the isochrone fitting of the observed CMD, which provided us with updated values of the cluster age, distance, average color excess, and metallicity. We estimated an absolute age of $13.1 \pm 0.5$ Gyr, consistent with the age of 47~Tucanae, an average color excess $E(B-V) = 0.64\pm 0.01$, and a true distance modulus $(m-M)_0=15.27 \pm 0.03$ that sets the cluster distance at 11.3 kpc from the Sun. In addition, the photometric estimate of the cluster metallicity suggests [Fe/H]$\approx-0.9$, which is $\sim 0.2$ dex smaller than that of 47~Tucanae. We also determined the gravitational center and the density profile of the system from resolved stars. The latter is well reproduced by a King model. Our results confirm that NGC 6316 is another extremely old relic of the assembly history of the Galaxy.
\end{abstract}

\keywords{Globular clusters: individual (NGC 6316)--}


\section{Introduction} \label{sec:intro}
Globular Clusters (GCs) are ancient stellar systems that are ubiquitous in the Milky Way. As such, the study of their structural, kinematic and dynamical properties is of paramount importance if we are to comprehend their connection with the early stages of the Galaxy assembly \citep[see][for a detailed discussion]{Forbes2018}. In particular, a comprehensive study and characterization of GCs located within the Galactic bulge has been proven to be mandatory to trace the properties of the bulge stellar population in terms of kinematics, chemical abundances, and age \citep[see][]{bica06,Valenti2007,barbuy18,Pallanca2019,Ferraro2009,Ferraro2016,Ferraro2021}.  However, these studies present significant challenges since bulge GCs are very dense and distant, and their light is severely absorbed by the presence of dark clouds of dust and gas along the line of sight. Also, their structure can be subjected to strong distortions  due to the effects of tidal forces exerted by the bulge \citep{Nordquist1999, Chun2015}. As a consequence, their main parameters are still poorly constrained, although they seem to share very old ages between $\sim12$ and $\sim13$ Gyr \citep[e.g.][]{Kerber2018,Kerber2019,Ortolani2019,Cadelano2020,Ferraro2021}, and a wide range of metallicities $-1.6 \lesssim$ [Fe/H] $\lesssim -0.2$ \citep[e.g.][]{Valenti2007,Valenti2010}. They also seem to follow an age-metallicity correlation with the younger GCs being more metal-rich than the older GCs \citep[see][]{Saracino2019,Pallanca2021b}.

This study is part of a large ongoing program aimed at characterizing the GCs located in the innermost regions of the Galactic bulge (see, e.g., \citealt{Valenti2010, Lanzoni2007, Lanzoni2010, Ferraro2009,Origlia2011, Origlia2013, Massari2014, Ferraro2016, Saracino2016,
  Cadelano2018, Saracino2019, Pallanca2019, Pallanca2021b, Pallanca2021a, Ferraro2021}), which can provide us with insights regarding the processes that led to the formation of the central region of the Milky Way \citep{Lee2018}. This
paper is focused on the case of NGC 6316, which is a relatively scarcely investigated cluster, with poorly known and still debated properties.  For instance, the Harris catalog \citep[][2010 edition]{Harris1996} quotes [Fe/H]$=-0.45$ for its metallicity,
although values ranging from $-0.36$ to $-0.87$ are found in the literature \citep{Carretta2009,Dias2016, Conroy2018}). The cluster is located at a distance of 11.6 kpc \citep{Valenti2007} from the Sun and
follows a highly eccentric orbit confined within the bulge, with a perigalactic distance of only 1.45 kpc \citep{Baumgardt2019}. Due to its location within the bulge, its stellar population is highly
contaminated by field stars and obscured by the presence of dust clouds along the line of sight \citep{Sandell1987}.  The first color-magnitude diagram (CMD) of NGC 6316 was published by \citet{Davidge1992} in the $V$ and $K$ filters with observations
performed at the 3.6 m Canada-France Hawaii Telescope (CFHT). The authors comment on the remarkable resemblance between NGC 6316 and 47 Tucanae (hereafter, 47 Tuc) based on the brightness widths of their
horizontal branches, and they suggest that NGC 6316 is slightly more metal-rich than 47 Tuc. Finally, an age of 13.8 $\pm$ 0.3 Gyr has been derived by \citet{Conroy2018}, on the basis of stellar population
models.
The goal of this work is to carry out the first high-resolution photometric study of the inner regions of NGC 6316 to characterize its stellar population properties. This paper is structured as follows. In Section~\ref{sec:photometry} we describe the photometric
analysis performed on the acquired data set. In Section~\ref{sec:diff_red} we describe the differential reddening correction.  In Section~\ref{sec:dist_red} we discuss the morphology of the CMD of NGC 6316 and its comparison to that of 47 Tuc, describing how we used it to obtain first estimates of the distance modulus and the color excess. In Section~\ref{sec:age} we discuss the fitting of the differential reddening corrected CMD by means of three sets of isochrones, which provides us with updated values of the cluster distance, color excess, age, and metallicity. In Section~\ref{sec:rad_prof} we determine the center of gravity of the cluster, its radial density profile and its structural parameters. Finally, in  Section~\ref{sec:summary} we summarize our results.

\section{observations and Data reductions} \label{sec:photometry}

The present work is based on a proprietary data set obtained with the Wide Field Camera 3 (WFC3) onboard the $HST$ (GO:15232, PI: Ferraro; \citealp{Ferraro2017}). The data set is composed of seven images ($6 \times30$ s and 1$\times$ 656 s exposures) in the F555W filter, and seven images ($6\times12$ s and $1 \times $ 643 s exposures) in the F814W filter. To avoid the inter-chip gap of the WFC3, in each pointing the cluster center was positioned in chip 1, while chip 2 samples distances out to $\sim$120$\arcsec$.

We have performed the photometric analysis using the standard package of \textrm{DAOPHOT IV} \citep{Stetson1987} on the -flc images, which are already corrected for dark-subtraction, flat-field, bias, and charge transfer efficiency. We have used a selection of 200 isolated stars in each image to accurately model the shape of the Point Spread Function (PSF). Its FWHM was set to 1.5 pixels ($\sim~0\arcsec.06$) and we used a 20-pixel radius ($\sim~0\arcsec.8$) to sample each of these isolated stars. Based on a $\chi^2$ test, in each image, the best PSF model was a Moffat function \citep{Moffat1969}. We thus applied it to all the sources detected above a 5$\sigma$ threshold from the local background level. We then built a master catalog containing the instrumental magnitudes and positions of each stellar source detected in at least 3 images. At the corresponding positions of all these sources, a fit was forced in each image using \textrm{DAOPHOT/ALLFRAME} \citep{Stetson1994}. The results have been combined together using \textrm{DAOPHOT/DAOMASTER}  to finally  obtain homogenized  magnitudes and their related photometric errors. 

The instrumental magnitudes have been calibrated to the VEGAMAG system using the zero points reported at the $HST$ WFC3 website\footnote{\url{https://www.stsci.edu/hst/instrumentation/wfc3/data-analysis/photometric-calibration/uvis-photometric-calibration}}, namely, ZP$_{\rm F555W1}$ = 25.735 and ZP$_{\rm F555W2}$ = 25.720 for stars the detected in chips 1 and 2 in the F555W filter, respectively and ZP$_{\rm F814W1}$ = 24.598 and ZP$_{\rm F814W2}$ = 24.574 for the corresponding F814W magnitudes. Finally, we applied independent aperture corrections for each chip and filter. 

The instrumental positions of the stars present in our images have been corrected for geometric distortions in both chips, according to the procedure described by \citet{Bellini2011}. They were also converted into the absolute system celestial coordinates (RA and Dec) through cross-correlation with the stars in common with the $Gaia$-DR3 catalog \citep{Gaia2016,Gaia2022}. 

\begin{figure*}[ht!]
\plotone{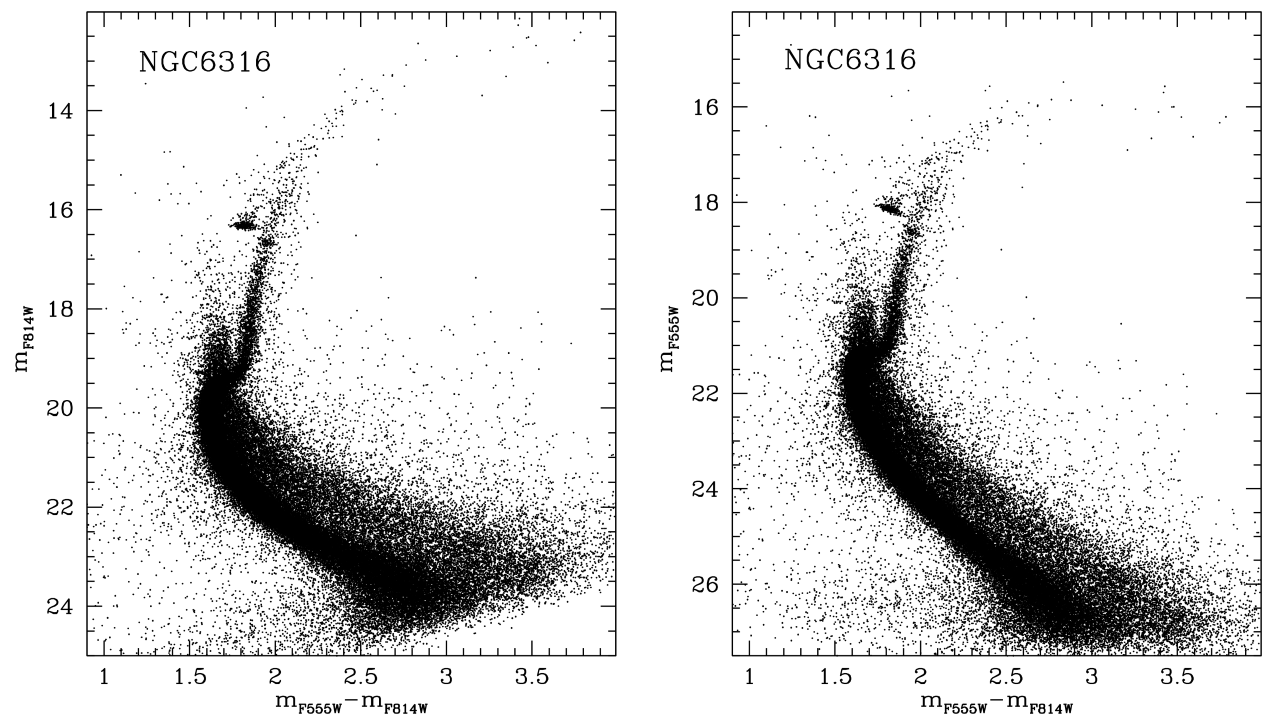}
\caption{CMD of NGC~6316 obtained from the $HST$-WFC3 data set used in this work, with the F814W and the F555W magnitudes plotted along the y-axis in the left and right panels, respectively. \label{fig:cmd_complete}}
\end{figure*}

We used this calibrated star catalog to create the first deep CMD of the cluster (Fig. \ref{fig:cmd_complete}). As mentioned in Section~\ref{sec:intro}, NGC 6316 is located within the bulge causing its CMD to suffer from a considerable amount of contamination from field interlopers. Nevertheless, its main features are easily distinguishable. The Main Sequence (MS) spans a range of magnitudes between 20.0 $<$ m$_{\rm F814W}$ $<$ 24.0 and shows the Turn-Off (TO) point at around m$_{\rm F814W}$ = 19.7. The CMD also shows a densely populated red Horizontal Branch (HB) at about m$_{\rm F814W}$ = 16.3 with no blue extension, a well-defined Sub Giant and Red Giant Branches (SGB and RGB, respectively) with the characteristic bump at m$_{\rm F814W}$ = 16.7, and the Asymptotic Giant Branch (AGB) above the HB. 

\section{Differential reddening correction} \label{sec:diff_red}

The catalog of GCs compiled by  \citet{Harris1996} quotes a reddening value of $E(B-V) = 0.54$ for NGC 6316. This is caused by dark clouds obstructing our line of sight, as already suggested by \citet{Sandell1987} and  \citet{Heitsch1999} who  showed that there is a small but non-negligible differential reddening in the direction of the cluster.

The presence of differential reddening has the effect of broadening the main evolutionary sequences on the CMD, and needs to be accounted for in order to accurately derive the cluster properties. To correct for this, we made use of the iterative method fully described in  \citet{Pallanca2019,Pallanca2021a} and \citet{Cadelano2020} that we briefly summarize in the following. We begin by selecting a reference sample of stars located at $r < 30\arcsec$ from the center of the cluster in order to avoid field interlopers on the CMD, and between magnitudes 16.0 $<$ m$_{\rm F555W}$ $<$ 24.0, i.e., along the MS, SGB, and RGB. The CMD created from the reference sample was divided vertically into magnitude bins of 0.5 mag, except at 20.0 $<$ m$_{\rm F555W}$ $<$ 24.5, where we used 0.25 magnitude bins to obtain a finer sampling. We then estimated the sigma-clipped median values of the (m$_{\rm F555W}$ $-$ m$_{\rm F814W}$) color and the m$_{\rm F555W}$ magnitude of each bin. By interpolating these medians we created a Mean Ridge Line (MRL), which we used to estimate the distance $\Delta X$ along the direction of the reddening vector to all the stellar sources in the reference sample. The reddening vector is defined using the extinction coefficients $R_{\rm F555W}$ = 3.227 and $R_{\rm F814W}$ = 1.856, appropriate for TO stars ($\sim$G2V type) and obtained from \citet{Cardelli1989} and \citet{Girardi2002}, under the assumption of the standard extinction coefficient $R_{V}=3.1$\footnote{Several authors have previously suggested that the extinction law in the direction of the Galactic bulge is variable, depending on the line of sight (see \citep{Stasinska1992}, \citep{Udalski2003}) with values between 2.5 (\citep{Nataf2013}; \citep{Pallanca2021a,Pallanca2021b}) to 3.2 \citep{bica06}. However, since we lack the near-infrared observations necessary to estimate the value of $R_{V}$ in the direction of NGC~6316,  we have adopted the standard extinction coefficient $R_{V}=3.1$.}. Finally, we assigned a $\Delta X$ value to each source in our catalog as the $\sigma$-clipped median of the $\Delta X$ values measured for the $n$ closest reference stars. Each resulting value of $\Delta X$ can be then transformed into the relative differential reddening $\delta E(B-V)$ as:

\begin{equation}
    \delta E(B-V) = \frac{\Delta X}{\sqrt{R^{\rm 2}_{\rm F555W} + (R_{\rm F555W}-R_{\rm F814W})^{\rm 2}}}
\end{equation}

\begin{figure*}[ht!]
\plotone{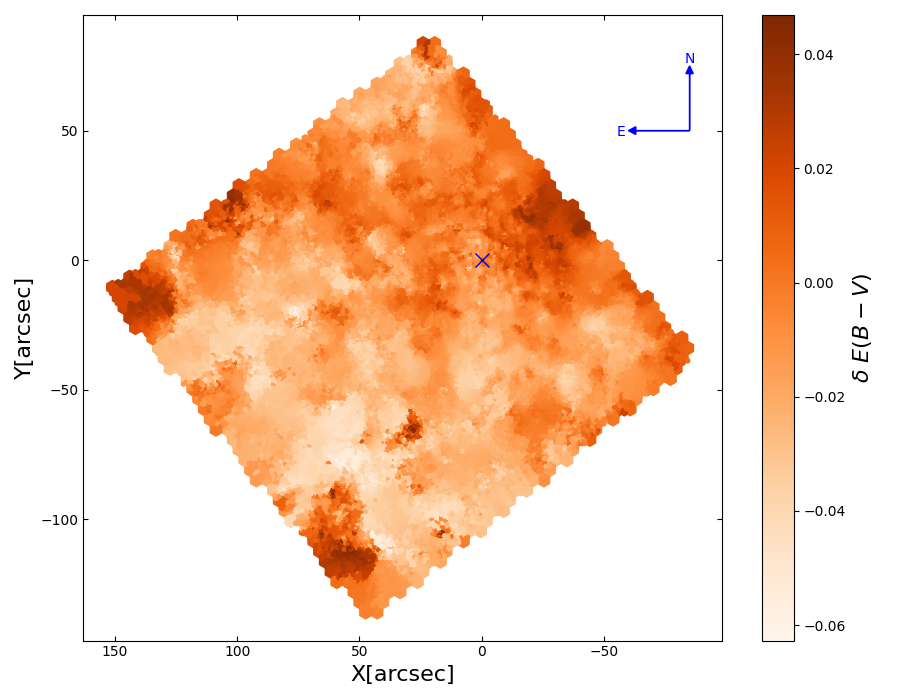}
\caption{Differential reddening map of NGC 6316 within the field of view of the WFC3. The colorbar on the right codifies the severity of the relative differential reddening. Lighter scale represents less extincted areas while darker scales represent more extincted areas. The coordinates along the x- and y-axes are reported with respect to the cluster center (blue cross).}
\label{fig:redmap}
\end{figure*}

\begin{figure*}[ht!]
\plotone{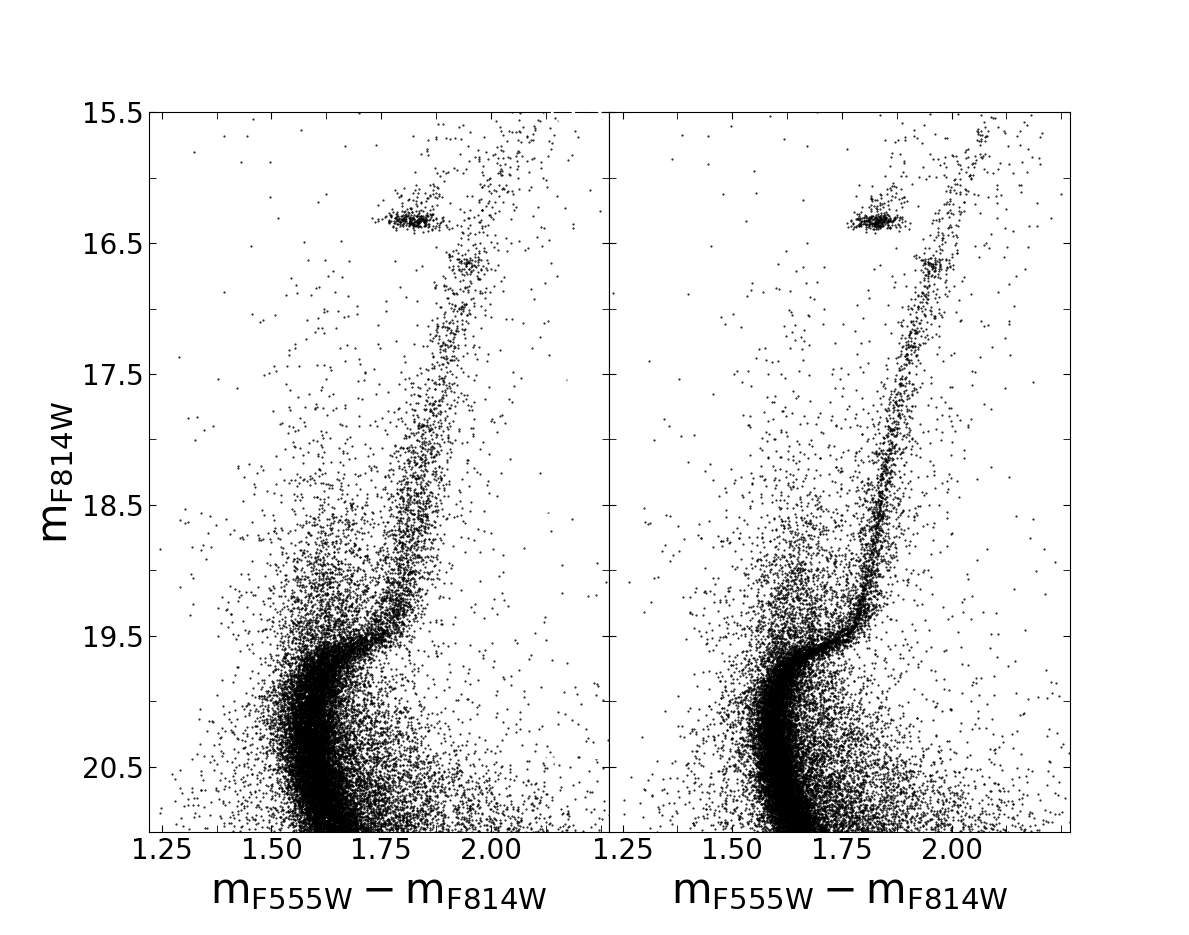}
\caption{The left panel shows the original CMD obtained from the $HST$ WFC3 used in this work. The right panel shows the same CMD after it has been corrected for differential reddening.}
\label{fig:cmd_debv}
\end{figure*}

We performed this iteration four times using the 50, 30, 25, and 15 closest stars to increase the spatial resolution in each iteration. The resulting reddening map is presented in Fig. \ref{fig:redmap}. It clearly shows the inhomogeneity of the medium causing the differential reddening across the cluster with a patchy structure and, possibly, a slightly larger average extinction in the Northern hemisphere. We found the differential reddening of the cluster to vary between $-0.063$ $<$ $\delta E(B-V)$ $<$ 0.047, within the sampled field of view. 
To make more evident the improvement in the definition of the evolutionary sequences in the CMD after the application of the differential reddening correction, in Fig. \ref{fig:cmd_debv} we show a comparison between the original and the corrected CMDs. After this correction, the main features of the CMD such as the MS, the SGB, the RGB and the HB look significantly sharper. 

\begin{figure*}[ht!]
\plotone{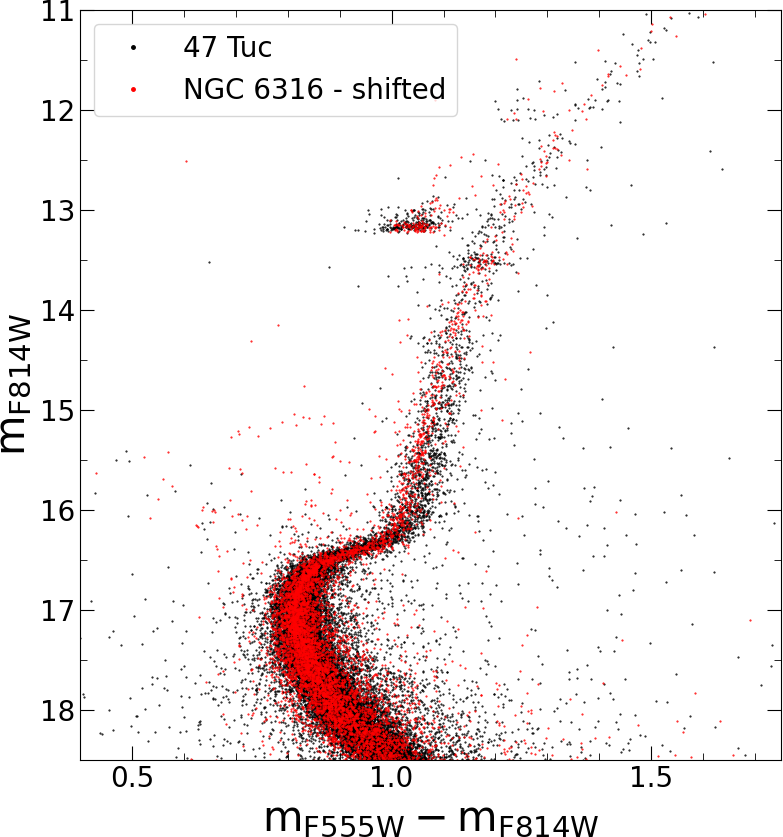}
\caption{Comparison between the CMD of 47 Tuc (black dots) and that of
  NGC 6316 properly shifted in magnitude and color to match the former
  (see Section~\ref{sec:age}).  For a cleaner visualization, only
  stars located at $r< 20\arcsec$ from their respective centers and
  with high-quality photometry (i.e., a sharpness parameter
  $|sh|\leq0.05$) are plotted in both the CMDs.}
\label{fig:cmd_overlap}
\end{figure*}

\begin{figure*}[ht!]
\plotone{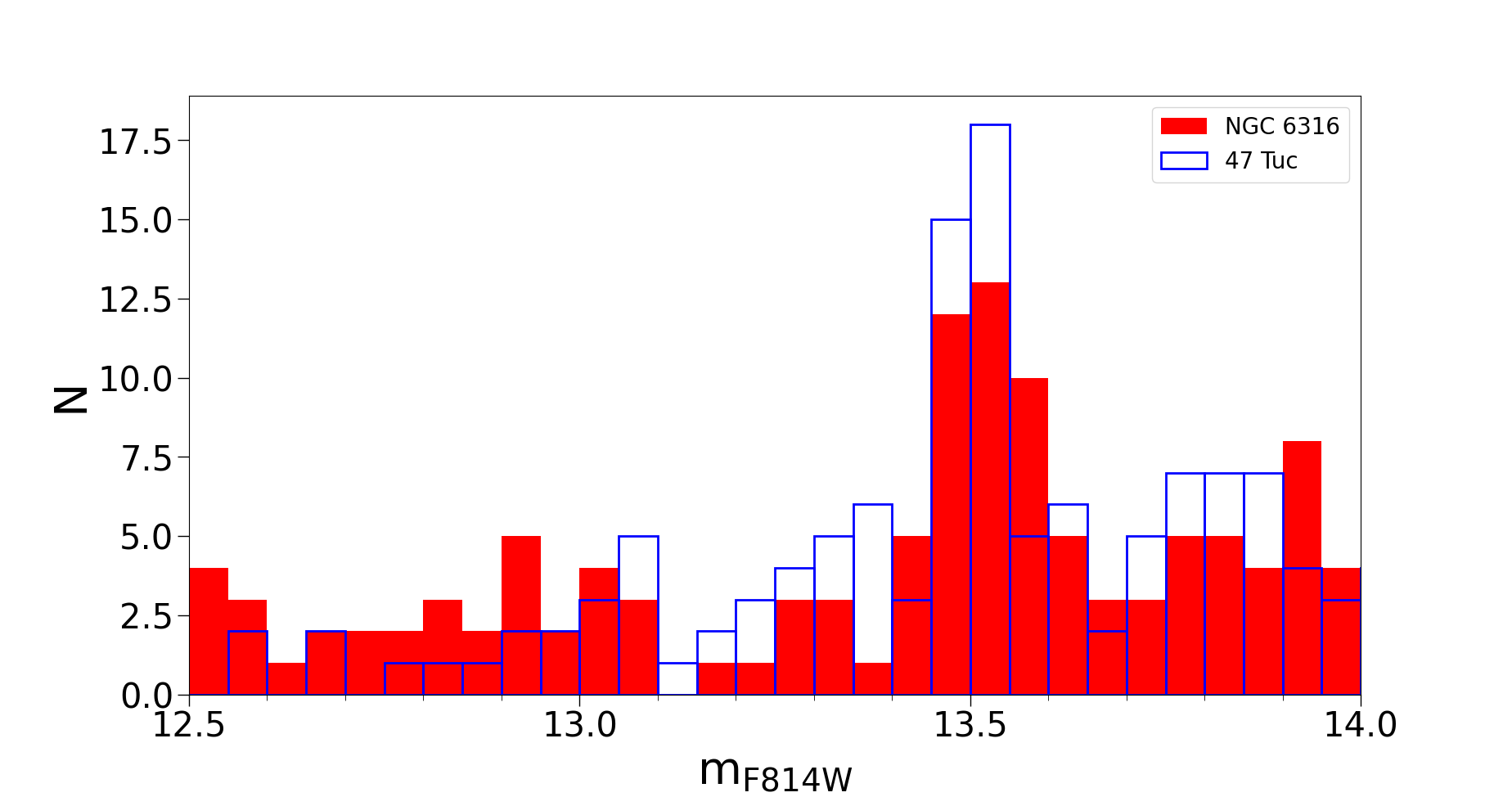}
\caption{Luminosity functions of the RGB stars observed in 47 Tuc (blue histogram) and in NGC 6316 (red histogram, after the shift applied to match the CMD of 47 Tuc), with the highest peaks corresponding to the locations of the RGB bump (m$_{\rm F814W}$ = 13.5) in the two systems.} 
\label{fig:histogram}
\end{figure*}

\section{Comparison with 47 Tuc} \label{sec:dist_red}

Since 47 Tuc is one of the most studied GCs in the Milky Way and its parameters have been very well constrained by several independent studies, we can use its CMD as a template to perform a detailed comparison with the CMD of NGC~6316 corrected for differential reddening. This will provide a first empirical hint on its basic properties, as the overall morphology of the evolutionary sequences closely resemble those observed in 47 Tuc \citep{Davidge1992} suggesting similar metallicities and ages ([Fe/H] = -0.76 and t = 12 - 13 Gyr for 47 Tuc; e.g., \citet{Carretta2009}, \citet{Thompson2020}, and references therein). To do this, we have used 4 images of 47 Tuc in the F555W and F814W filters (30 s and 0.5 s, per filter) acquired with the WFC3 through the GO Proposal 11664 (PI: Thomas Brown). For the photometric analysis, we have followed the exact same procedure described in Section~\ref{sec:photometry}, thus obtaining a CMD in the same photometric bands of NGC~6316.
This allowed us to perform a direct comparison of the two CMDs. To best highlight the evolutionary sequence populated by likely cluster members, in the comparison we considered only stars populating the inner portion of both clusters ( i.e., stars at r $<$ 20$\arcsec$ from their respective centers) and with a high-quality photometry ($-$0.05$<$sharpness$<$0.05). 
We found that shifts in magnitude and color equal to $\Delta$m$_{\rm F814W} = -3.16$ and $\Delta$(m$_{\rm F555W} -$ m$_{\rm F814W}) = -0.78$, respectively, are required to make the sequences of NGC~6316 matching those of 47 Tuc. The result is shown in Fig. \ref{fig:cmd_overlap}: indeed, the resemblance between the two CMDs is remarkable. In fact, the level and the morphology of both HBs is basically the same, as well as the luminosity and morphology of the TO and SGB regions, thus suggesting that the two
clusters are essentially coeval.  Moreover, the RGB bump is located exactly at the same luminosity in both clusters (m$_{\rm F814W}$ $\sim$ 13.5 in the figure). This is further confirmed by the comparison between the RGB luminosity functions in the magnitude range 12.5 $<$ m$_{\rm F814W}$ $<$ 14.0 (Fig.~\ref{fig:histogram}), where the highest peaks correspond to the RGB bumps in the two systems. On the other hand, it is worth noticing that the RGB branch of NGC~6316 is slightly but systematically bluer than that of 47 Tuc, possibly suggesting a small difference in terms of $[\rm Fe/H]$, with NGC~6316 being slightly metal poorer than 47 Tuc. 
The relative shifts needed to match the evolutionary sequences of the two clusters also provided us with first estimates of NGC~6316 distance and average color excess. By adopting the coefficients $R_{\rm F555W}$ and $R_{\rm F814W}$ from Section~\ref{sec:diff_red}
and assuming for 47~Tuc a color excess of 0.03 $\pm$ 0.01 and a distance modulus of 13.21 $\pm$ 0.06 (random, $\pm$ 0.03 systematic; \citealt{Brogaard2017}), NGC~6316 turns out to have an average color excess $E(B-V)=0.61\pm0.03$ and a distance modulus
$(m-M)_{0}=15.32\pm0.06$.

\section{Isochrone fitting}
\label{sec:age}
In order to confirm these preliminary hints, we have applied a Bayesian procedure similar to that used by \citet[][see also   \citealp{Saracino2019,Cadelano2019,cadelano20psr}]{Cadelano2020} that allows to derive a ``photometric" determination of the age, distance, color excess and metallicity performing an isochrone fitting
of the observed CMD. To this end, the CMD of NGC~6316 has been compared to a grid of different isochrone sets computed in a suitable range of different ages and metallicities, distance moduli and color
excesses. We extracted the isochrones from three different databases, namely, DSED \citep{Dotter2008}, BaSTI \citep{Pietrinferni2021}, and
PARSEC \citep{Marigo2017}. For each isochrone, we assumed a standard He content $Y$ = 0.25 and $[\alpha \rm /Fe]$ = +0.4, as typically measured for bulge GCs.  Following the same computational approach described in \citet[][see their Section 4.2]{Cadelano2020}, we
compared the observed CMD of the cluster with each family of stellar models adopting a Markov Chain Monte Carlo (MCMC) sampling technique. In doing this, we assumed a Gaussian likelihood function (see equations 2 and 3 in \citealt{Cadelano2020}).  To sample the
posterior probability distribution in the n-dimensional parameter space, we used the \texttt{emcee} code \citep{foreman13,foreman19}. We explored a wide range of ages from 10 to 15 Gyr, in steps of 0.2 Gyr,
assuming a flat prior within this range. As quoted in the Introduction, the metallicity of this cluster is not well defined through spectroscopy, since literature works show a large spread in the derived [Fe/H] values (see Section~\ref{sec:intro} and \ref{sec:diff_red}). Therefore, we explored the metallicity space
using a flat prior in a range from $[\rm Fe/H]=-1.2$ to $[\rm Fe/H]=-0.2$, in steps of 0.05 dex. The isochrones absolute magnitudes were converted to the observed frame using color excesses and distance moduli following Gaussian prior distributions peaked at
the previously determined values, i.e., $E(B-V)=0.61 \pm 0.03$ and $(m-M)_0=15.32 \pm 0.06$, respectively (see
Section~\ref{sec:diff_red}). We used temperature-dependent extinction coefficients from \citet{Casagrande2014}.  In order to minimize the
contamination from field interlopers and maximize the accuracy of the result, we performed the fit only on stars within $30\arcsec$ from the center (see Section~\ref{sec:rad_prof}), with a high-quality
photometry (i.e. sharpness parameter $|sh|\leq0.025$) and in the magnitude range $21<\rm m_{\rm F814W}<17$, corresponding to the CMD region most sensitive to stellar age and metallicity\footnote{In the central regions of this bulge GC, the $Gaia$-DR3 data \citep{Gaia2016,Gaia2022} sample only a small fraction of bright stars, thus not substantially improving in the selection of cluster members for building the necessary CMDs.}.

\begin{figure*}[ht!]
\begin{center}
\includegraphics[scale=0.4]{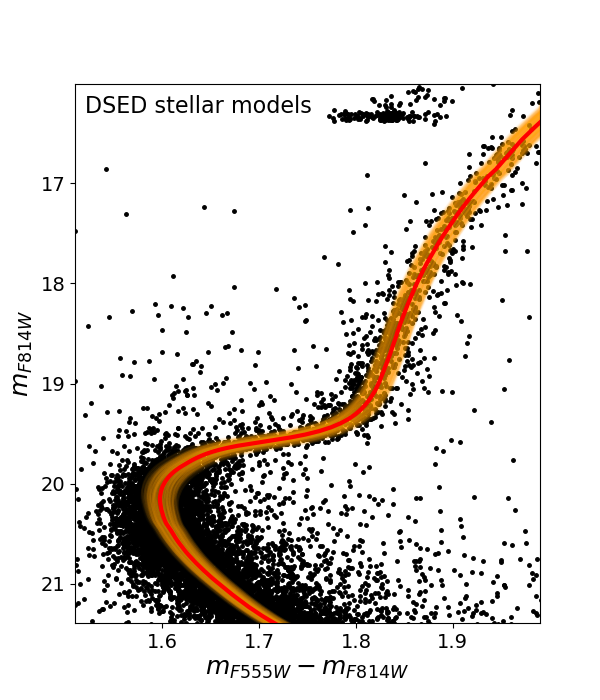}
\includegraphics[scale=0.25]{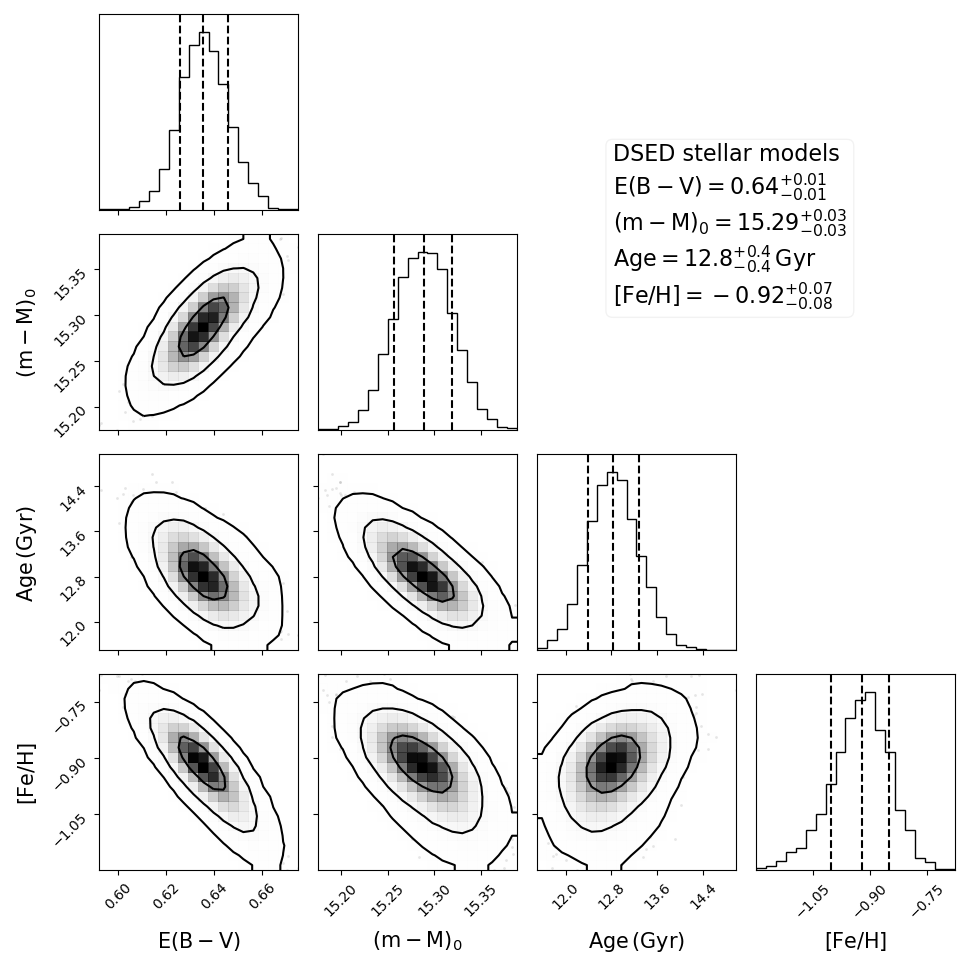}

\includegraphics[scale=0.4]{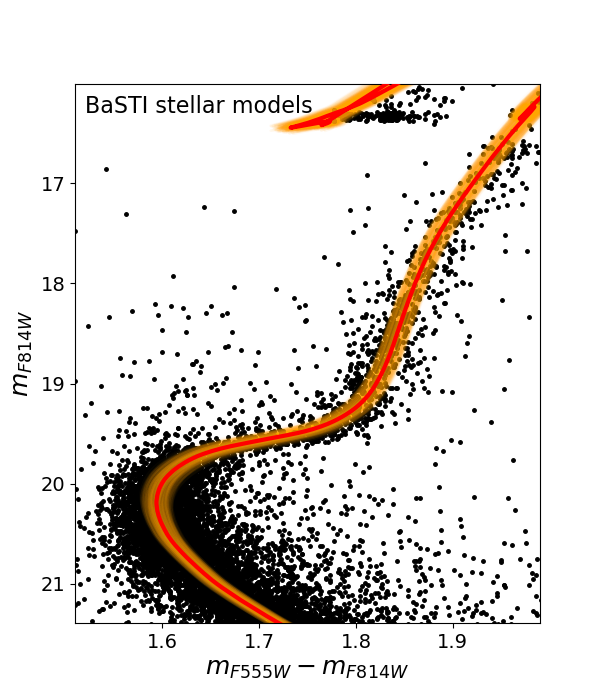}
\includegraphics[scale=0.25]{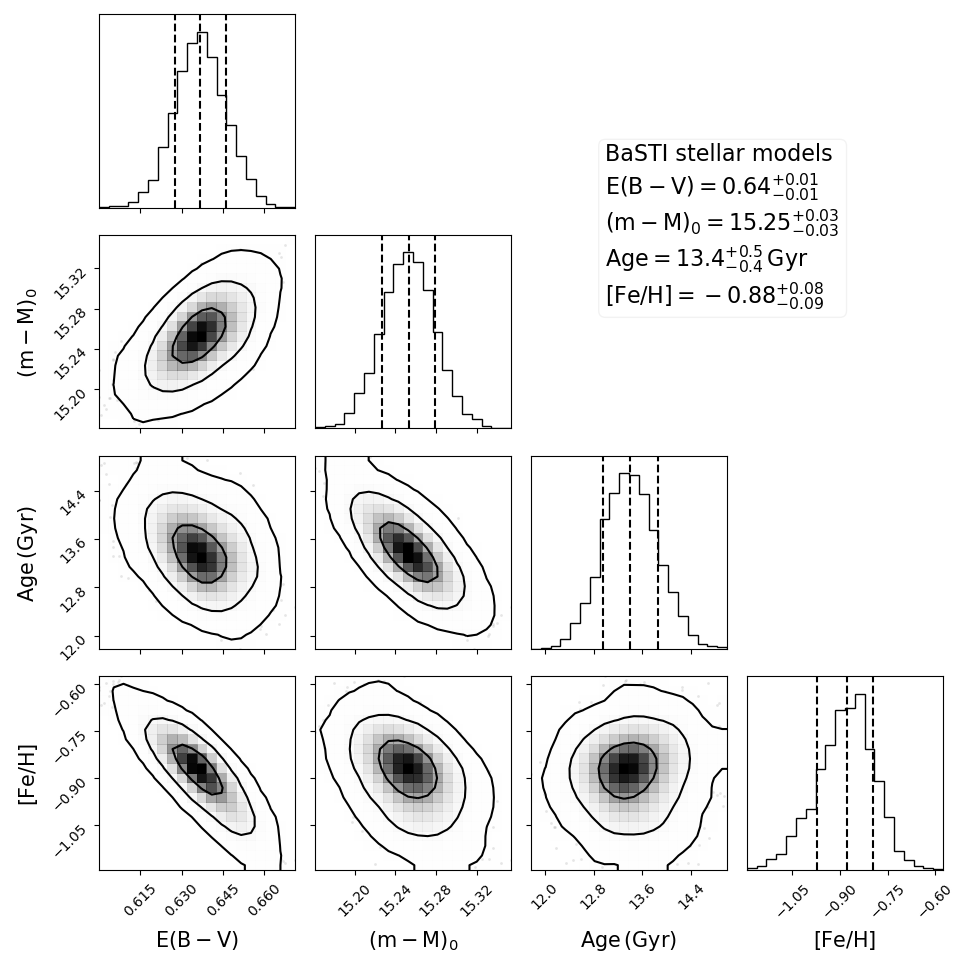}

\includegraphics[scale=0.4]{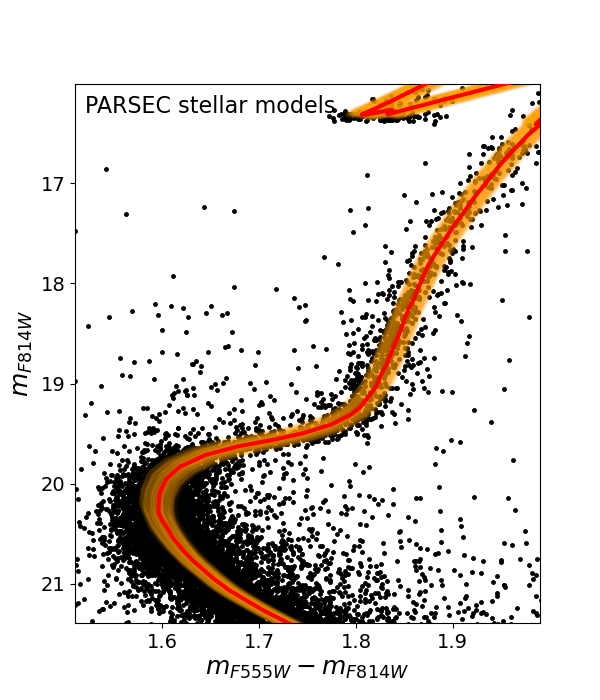}
\includegraphics[scale=0.25]{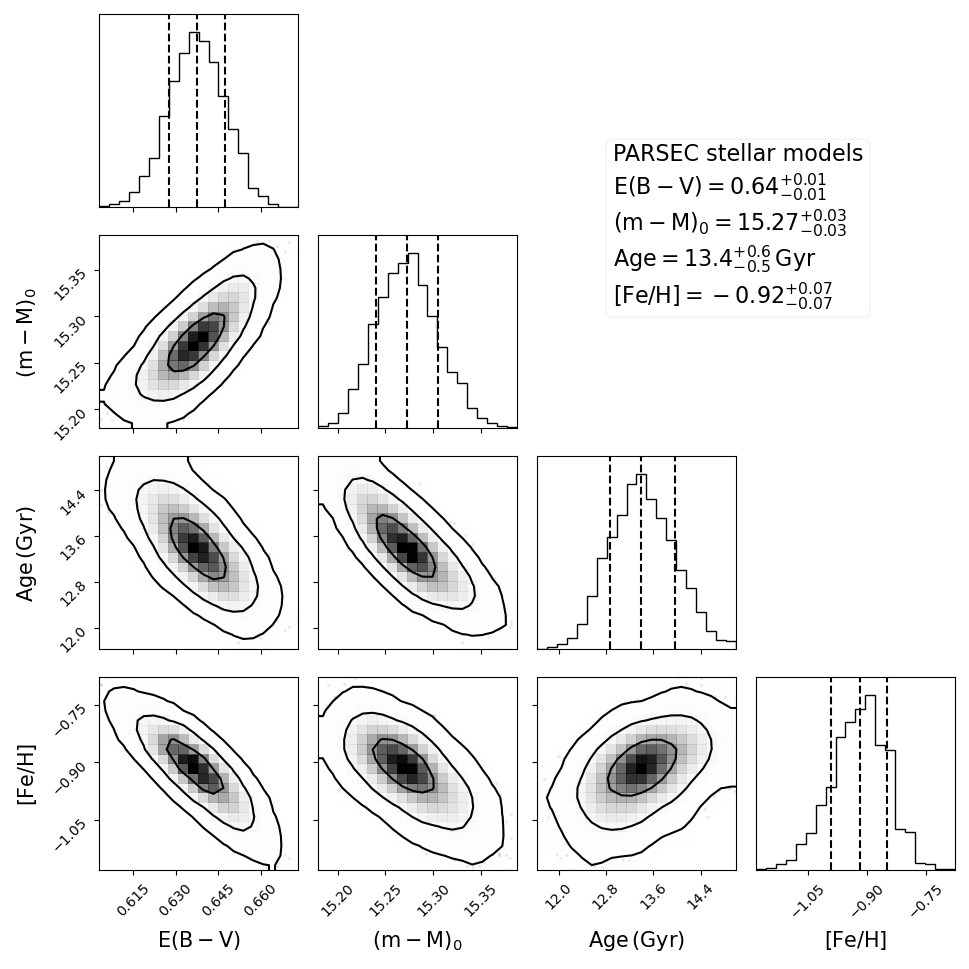}

\caption{{\it Left panels}: CMD of NGC 6316 with the best-fit DSED,   BaSTI and PARSEC isochrone (top, middle and bottom panels,   respectively) plotted as a red solid line and the $1\sigma$ envelope shaded in orange. {\it Right panels}: Corner plots showing the one- and two-dimensional projections of the posterior probability   distributions for all the parameters, as obtained from DSED, BaSTI   and PARSEC isochrone (top, middle and bottom panels, respectively). The contours correspond to the $1\sigma$, $2\sigma$ and $3\sigma$ levels.}
\label{fig:isochrones}
\end{center}
\end{figure*}

The results obtained in terms of age, metallicity, distance modulus and color excess, are shown in Figure~\ref{fig:isochrones} for the three adopted sets of theoretical models. The left-hand panels show
the CMD and the best-fit isochrones.  The one- and two-dimensional posterior probabilities for all of the parameter combinations are presented in the right-hand panels as corner plots. The best-fit values and their uncertainties (based on the $16^{th}$, $50^{th}$,
$84^{th}$ percentiles) are also summarized in Table \ref{tab:1}. As apparent, the resulting values of color excess and distance   modulus are in good agreement with those obtained from the direct   comparison with the CMD of 47 Tuc. We also verified that the results
  remain unchanged within the uncertanties if a uniform prior spanning   of a larger interval of values for both the $E(B-V)$ and $(m-M)_0$ is assumed.  The best-fit age turns out to be 12.8 Gyr if the DSED models are assumed, and 13.4 Gyr with the two other sets of
  isochrones. These results are in agreement within the uncertainties and the differences can be explained by the fact that each model uses slightly different values for the same parameters (like, the solar abundances, reaction rates, electron conduction opacities, mixing length, etc...).
Finally, as expected from the CMD comparison, the isochrone fitting
results suggest that NGC 6316 hosts a stellar population slightly
metal poorer than that of 47~Tuc, with [Fe/H]$< -0.88$.

\begin{table}[ht!]
	\begin{center}
		\caption{Best-fit parameter values for DSED, BaSTI and PARSEC models.}\label{tab:1}
		\begin{tabular}{|c|c|c|c|c|} 
			\hline
			Model	&   Age    & $[\rm Fe/H]$ & $E(B-V)$   & $(m-M)_{0}$ \\
				& [Gyr]   &   dex &    [mag]    &  [mag]     \\			
			\hline
			DSED   & $12.8\pm0.4$ & $-0.92^{+0.07}_{-0.08}$ &  $0.64\pm0.01$ & $15.29\pm0.03$ \\ 
			BaSTI & $13.4^{+0.5}_{-0.4}$ & $-0.88^{+0.08}_{-0.09}$ &  $0.64\pm0.01$ & $15.25\pm0.03$ \\ 
			PARSEC & $13.4^{+0.6}_{-0.5}$ & $-0.92\pm0.07$ &  $0.64\pm0.01$ & $15.27\pm0.03$ \\ 
			\hline
                        average & $13.1\pm 0.5$ & $-0.9 \pm 0.1$ & $0.64 \pm 0.01$ & $15.27\pm 0.03$\\
                        \hline
		\end{tabular}
	\end{center}
\end{table}

\section{the density profile of ngc 6316} \label{sec:rad_prof}

\subsection{Center of gravity}

In order to derive the structural parameters of NGC 6316, we need a reliable measurement of the cluster gravitational center ($C_{\rm grav}$). To this aim, we applied a procedure that, starting from a first-guess value, iteratively computes the average of the projected x- and y-positions of a selected sample of stars, and reaches convergence when the difference between two consecutive iterations is negligible (see, e.g., \citealp{Montegriffo1995,Lanzoni2007,Lanzoni2019,Raso2020}). As a first-guess center, we adopted the coordinates reported in the catalog of Orbital Parameters of Galactic Globular Clusters \citep{Baumgardt2018}. The sample of stars used in the computation has been selected adopting different magnitude limits (20.5 $<$ m$_{\rm F555W}$ $<$ 22.3 in steps of 0.5 mag) and different distances from the cluster center (from 15$\arcsec$ to 35$\arcsec$ in steps of 5$\arcsec$). In order to mitigate the contamination due to field interlopers, we considered only stars with colors between 1.5 $<$ m$_{\rm F555W}$-m$_{\rm F814W}$ $<$ 2.0. The final position of $C_{\rm grav}$ is the average of the centers determined for each combination of magnitude and distance limits: R.A. = 17$^{\rm h}$ 16$^{\rm m}$ 37$^{\rm s}$.2  and Dec = -28$^{\circ}$ 08$\arcmin$ 23$\arcsec$.42 with an uncertainty of 0.2$\arcsec$. Our estimate is located $\sim$ 1.6$\arcsec$ north-west from the initial first-guess center reported by \citet{Baumgardt2018}. 

\subsection{Stellar density profile}

The WFC3 field of view covers a region of $162\arcsec \times 162\arcsec$ and thus only partially samples the cluster radial extension. Therefore, in order to determine the entire density profile we need a complementary wide-field catalog of stars, extending up to the outskirts of the cluster and beyond. We used a star catalog covering a circular area around NGC~6316 with a $1^{\circ}$ radius extracted from $Gaia$-DR3. The $Gaia$ magnitudes have been corrected for the effects of differential reddening using the publicly available $E(B-V)$ maps by \citet{schlafly11}. These maps show strong reddening variations within the field of view sampled by the adopted $Gaia$ catalog, with $E(B-V)$ ranging from 0.25 to 2.05. Such a correction is therefore mandatory since, as demonstrated by \citet{Cadelano2017}, reddening variations across the wide-fields used to sample the cluster external regions can lead to severe deviations from the intrinsic density profile of the system. 

\begin{figure*}%
    \centering
    \includegraphics[scale=0.45]{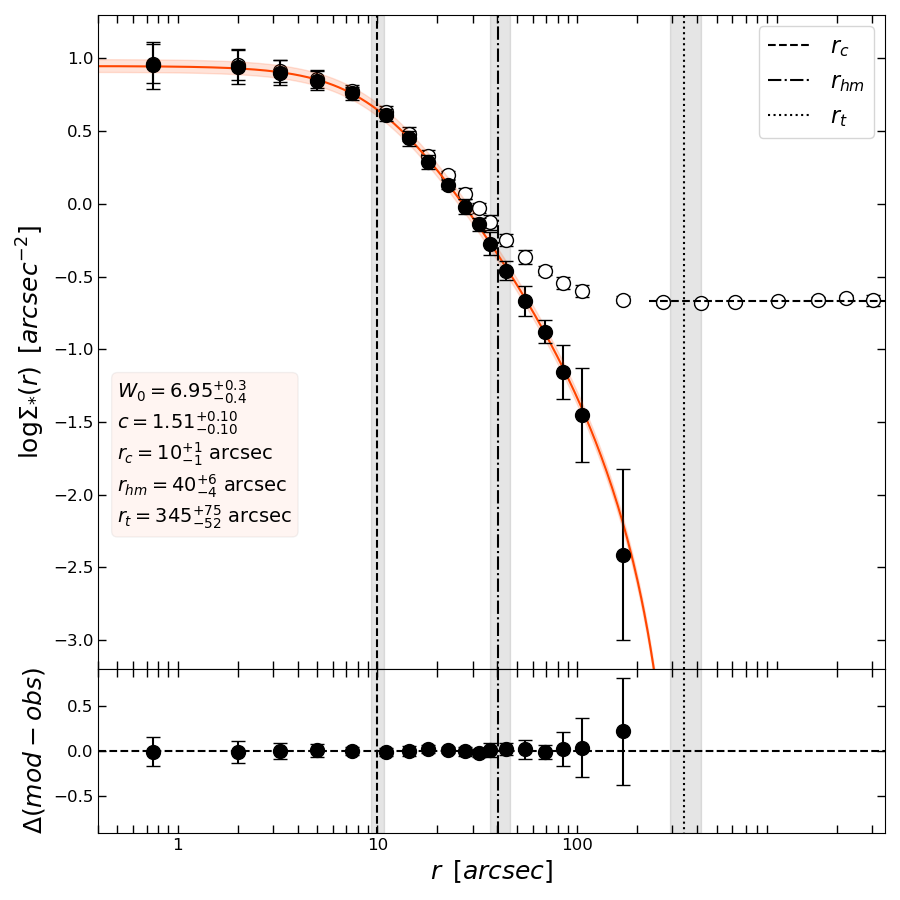} %
    \caption{Observed (empty circles) and background-subtracted (filled circles) density profile of NGC 6316. The dashed horizontal line is the background density value used to decontaminate the cluster profile. The solid red curve is the best-fit King model to the cluster density profile and the red stripe marks the envelope of the $\pm~ 1\sigma$ solutions. The dashed, dot-dashed and dotted vertical lines mark the best-fit cluster's core, half-mass and tidal radii, respectively, and their corresponding $1\sigma$ uncertainties are represented with the gray stripes. The best-fit values of some structural parameters are also labeled. The bottom panel shows the residuals between the best-fit King model and the cluster density profile.}
    \label{fig:density_profile}%
\end{figure*}

To estimate the density profile, we used the procedure described, e.g., in \citet{Miocchi2013,Lanzoni2019,Raso2020}. Briefly, in both the catalogs, we selected the stars brighter than the MS-TO applying a cut parallel to the reddening vector in order to mitigate possible residual effects of differential reddening. Such a selection naturally includes only stars with approximately the same mass. For the WFC3 data set we selected 17 concentric rings around $C_{\rm grav}$ with radii up to $120\arcsec$. This allowed us to sample the density profile of the cluster inner regions. In the case of the $Gaia$-DR3 data, we selected 12 concentric rings with radii between 60$\arcsec$ and 3600$\arcsec$ in order to sample the outermost regions of the cluster, as well as the Galactic background density. Each ring is divided into four subsectors and the average and standard deviation of the different density measurements in each sub-sector have been chosen as the resulting density value of the ring and its related uncertainty. The $Gaia$ profile was then vertically rescaled to match that of the WFC3  using 4 common points in the radial range between 60$\arcsec$ and 120$\arcsec$. This provided us with a full stellar density profile of the cluster from the central region and well beyond its outskirts (see the empty circles in Fig.~\ref{fig:density_profile}). The profile shows a flat behavior out to $\sim~5\arcsec$, indicating that this is likely not a post core-collapsed system, then it steadily decreases out to 200$\arcsec$ from the center. At distances larger than $r=200\arcsec$ the stellar background becomes dominant over the cluster population. This is clearly illustrated by the well-defined plateau present in the outermost portion of the density profile. Indeed, the spatial distribution of background stars is expected to be approximately uniform on the considered radial scale. In order to obtain the intrinsic density profile of the system, the background contribution must be removed from the observed profile. 

For this reason, the level of Galactic field contamination has been estimated by averaging the seven outermost points aligned in the plateau (see the horizontal dashed line in Fig. \ref{fig:density_profile}), and then subtracted from the observed distribution (empty circles in Fig. \ref{fig:density_profile}),  thus obtaining the background-decontaminated star density profile of NGC~6316 (filled circles in Fig. \ref{fig:density_profile}). It is apparent that after the field subtraction, the profile remains almost unchanged at small radii, which are dominated by the cluster population, while it significantly decreases in the most external regions, where it turns out to be below the Galactic background. This clearly indicates that an accurate measurement of the background level is crucial for the reliable determination of the outermost portion of the density profile.

The cluster’s structural parameters have been derived by fitting the background-decontaminated profile with a single-mass King model  \citep{King1966}, assuming spherical symmetry and orbital isotropy. Following \citet{Raso2020}, we performed the fit using a MCMC approach implemented by the {\textrm emcee} package \citep{foreman13,foreman19}. We assumed uniform priors on the parameters of the fit (i.e. the King concentration parameter $c$, the core radius $r_c$, and the value of the central density). Therefore, the posterior probability distribution functions are proportional to the likelihood $L = \exp(-\chi^2/2)$, where the $\chi^2$ statistic is calculated between the measured density values and those predicted by the whole family of adopted models.
The best-fit model is shown as a red line in Fig.~\ref{fig:density_profile}, and the resulting structural parameters are also labeled. The density profile is nicely reproduced by a King model with an intermediate concentration of about $c=1.51$, and a small core and tidal radii of about $10\arcsec$ and $350\arcsec$, respectively. As shown by the residuals of the fit (bottom panel of Fig.~\ref{fig:density_profile}), no signs of deviations from the King distribution are observed neither in the internal, nor in the external regions, thus confirming that the cluster is not in a core-collapse phase and it is not subjected to a severe tidal stripping in the outer regions due to its motion in the bulge potential field. 

For the sake of completeness, we have also estimated the central and half-mass relaxation times as log($t_{rc}$/yr) = 8.11 and log($t_{rh}$/yr) = 9.09, following Equation (10) and Equation (11) in \citet{Djorgovski1993}, respectively. The complete list of the basic parameters estimated for NGC 6316 is summarized in Table \ref{table:tab_params}.

\begin{figure*}%
    \centering
    \includegraphics[scale=0.65]{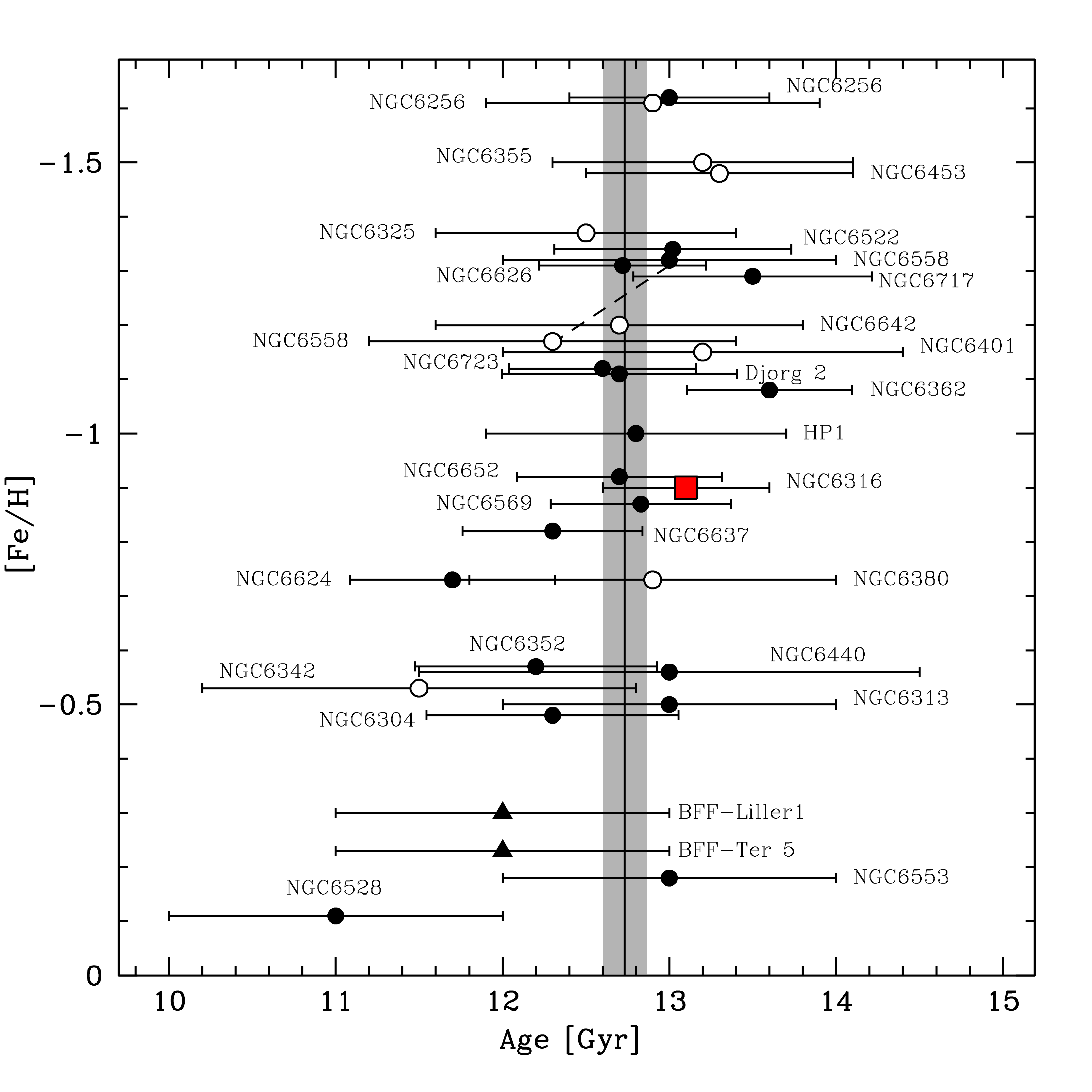} %
    \caption{Age-metallicity distribution of the bulge GCs with available age estimates. NGC 6316 is marked as a large red square. The solid circles mark the data from \citet[][see their Fig. 16]{Saracino2019}, and \citet[][see
        their Fig. 12]{Oliveira2020} and \citet{Cadelano2020}, while the empty circles are from \citet{Cohen2021}. Two clusters are in common between the \citet{Cohen2021} data set and the previous sample: NGC 6256, which has essentialy the same values in both catalogs, and NGC 6558, for which a lower metallicity and a smaller age have been estimated in the most recent study (the dashed line connects the two data points). Also plotted, are the age and metallicity of the old stellar populations in the two Bulge Fossil Fragments (Terzan 5 and Liller 1; triangles) discovered so far       \citep{Ferraro2009,Ferraro2016,Ferraro2021}. The gray vertical strip marks the weighted average and 1$\sigma$ uncertainty (12.7 $\pm$ 0.1 Gyr) of the entire sample.}
    \label{fig:fe_age}%
\end{figure*}

\section{summary and conclusions} \label{sec:summary}
 We used high-resolution $HST$ observations to characterize the   stellar population of the poorly studied GC NGC~6316 located in the   Galactic bulge. We obtained the first high-quality CMD of the system
  extending down to $\sim 3$ mag below its MS-TO, and we corrected it for the effects of differential reddening. The comparison between the CMD of NGC~6316 and that of 47~Tuc in the same filter combination showed a remarkable similarity between the respective
  stellar evolutionary sequences, apart from a slightly bluer color of the RGB in the former system, suggesting a slightly lower metallicity.  This comparison provided us with the first-guess values for the average color excess and the cluster distance, which we then adopted in the isochrone fitting of the CMD. This has been performed by using three different sets of models (DSED, BaSTI and
  PARSEC) and the weighted averages of the obtained results provided us with updated estimates of the cluster paramaters (see the last row of Table \ref{tab:1}). The average color excess in the direction of NGC 6316 turns out to be $E(B-V)=0.64 \pm 0.01$, which is in rather good agreement with the value of 0.61 estimated by \citet{Davidge1992}. The distance modulus $(m-M)_0 = 15.27 \pm 0.03$
  translates into a distance $d =11.3 \pm 0.3$ kpc from the Sun, which is in agreement with the values previously estimated by both \citet{Valenti2007}, and \citet[][$d = 11.152 ^{+0.393}_{-0.382}$
    kpc]{Baumgardt2021}. Note that the 
 \citet{Harris1996} catalog reports instead, a $\sim 1$ kpc shorter distance (10.4 kpc), and a 0.1 mag lower reddening (0.54). The best-fit age is $13.1 \pm 0.5$
  Gyr, which is in good agreement, as expected from the CMD comparison, with the age of 47~Tuc determined by different authors (e.g., $13.0 \pm 2.5$ Gyr by \citealp{Zoccali2001}; 12.75 $\pm$ 0.50
  Gyr by \citealp{Dotter2010}; 11.8 $\pm$ 1.6 Gyr by
  \citealp{Brogaard2017}; and 12.0 $\pm$ 0.5 Gyr by
  \citealp{Thompson2020}). Previous estimates of the age of NGC 6316 include 10.0 Gyr \citep{Santos2004} and 10.73 Gyr \citep{Zhang2010}, which are noticeably smaller than the ones obtained in this work.

Finally, all the three models suggest a metallicity below $-0.88$, with an average value [Fe/H]$= -0.9 \pm 0.1$. This is $\sim 0.2$ dex smaller than that measured for 47~Tuc \citep{Harris1996}, as expected from the comparison of the RGB color in the two clusters. Although accurate measurements of the cluster metallicity from high-resolution spectroscopy are needed to properly assess the metal content of NGC
6316, this result suggests that values as large as $\sim -0.4$ \citep[e.g.][]{Harris1996, Dias2016} are likely to be excluded.  

By assuming our photometric estimates, NGC 6316 well fits in the age-metallicity distribution drawn by the bulge GCs with available age estimates (see Fig. \ref{fig:fe_age}). Here we also included the two Bulge Fossil Fragments (Liller 1 and Terzan 5) identified so far
\citep{Ferraro2009,Ferraro2016,Ferraro2021}. These are peculiar stellar systems that, despite their appearance as genuine GCs, host multi-age stellar populations and could be the remnants of massive clumps that contributed to form the bulge at the epoch of the Galaxy
assembly. From the comparison, it is apparent that also these peculiar systems share very old ages with the
majority of GCs and Galactic field stars located in the bulge \citep{Zoccali2003,Clarkson2011,Valenti2013}.
The weighted mean age of the entire sample (27 GCs + 2 bulge fossil fragments) is $12.7 \pm 0.1$ Gyr. The values of age and metallicity estimated in this work, together with the orbital parameters (e.g., \citet{Baumgardt2019}) and results from the analysis of the integrals of motion \citep{Massari2019}, are consistent with NGC 6316 being formed $in~situ$ within the Galactic bulge. 

Finally, by combining the $HST$ observations with the wide-field, publicly available $Gaia$-DR3 catalog, we built the cluster stellar density profile from resolved stars.  This is very well reproduced by a King model with an intermediate concentration of about 1.51, a
compact core of $\sim$ 0.55 pc ($10\arcsec$) and an effective radius (i.e., the radius including half of the total light in projection) $r_e\sim 1.6$ pc ($29.8 \arcsec$). For comparison, the \citet{Harris1996} catalog quotes a larger concentration ($c=1.65$), the
same $r_c$, and a larger effective radius ($r_e=39\arcsec$).  Our updated estimates of the central and half-mass relaxation times are also slightly larger than those in \citet{Harris1996}.  The value of $t_{rc}$ ($\sim 7.4\times 10^7$ yr) compared with the cluster chronological age ($\sim 1.3\times 10^{10}$ yr) suggests that NGC~6316 is a dynamically old stellar system. Following \citet{Ferraro2012,
  Ferraro2018, Ferraro2019, Ferraro2020, Lanzoni2016} the level of central segregation of blue straggler stars (see their definition in \citealt{Sandage1953}) within a GC is a powerful empirical diagnostic
of the dynamical age of the host system, and it has been found to scale with the number of relaxation times suffered by the cluster since its formation. A future paper will be devoted to the accurate selection of the blue straggler population in NGC~6316, the determination of their radial distribution, and the measure of their level of central segregation.  We also plan to measure the values of the three dynamical indicators recently proposed by \citet{bhat22}. This will provide independent and empirical estimates
of the dynamical age of the cluster, and allow its comparison with that of the other GCs in the Galactic system.  
  
\begin{table}
\caption{Structural Parameters of NGC 6316.}
\begin{center}
\begin{tabular}{ l l}
\hline
\hline
Parameter & Estimated value \\
\hline
Center of gravity               & $\alpha_{\rm J2000}= 17^{\rm h} 16^{\rm m} 37.2^{\rm s}$ \\
                                & $\delta_{\rm J2000} = -28^{\circ} 08\arcmin 23.42\arcsec$\\
Dimensionless central potential & $W_0=6.95^{+0.30}_{-0.40}$\\
Concentration parameter         & $c=1.51^{+0.10}_{-0.10}$ \\
Core radius                     & $r_c=10^{+1}_{-1}$ arcsec \\
Half-mass radius                & $r_h=40^{+6}_{-4}$ arcsec \\
Effective radius                & $r_e=29.8\pm 5$ arcsec \\
Truncation radius               & $r_t=345^{+75}_{-52}$ arcsec\\
Central relaxation time & $\log(t_{rc}/{\rm yr}) = 8.11$\\
Half-mass relaxation time & $\log(t_{rh}/{\rm yr}) = 9.09$\\
\hline
\end{tabular}
\end{center}
\label{table:tab_params}
\end{table}

\begin{acknowledgments}
This research is part of the project Cosmic-Lab at the Physics and Astronomy Department of the University of Bologna (http://www.cosmic-lab.eu/Cosmic-Lab/Home.html). The research has been funded by project Light-on- Dark, granted by the Italian MIUR through contract PRIN-2017K7REXT (PI: Ferraro).

This work has made use of data from the European Space Agency (ESA) mission
{\it Gaia} (\url{https://www.cosmos.esa.int/gaia}), processed by the {\it Gaia}
Data Processing and Analysis Consortium (DPAC,
\url{https://www.cosmos.esa.int/web/gaia/dpac/consortium}). Funding for the DPAC
has been provided by national institutions, in particular the institutions
participating in the {\it Gaia} Multilateral Agreement.

\end{acknowledgments}

%

\vspace{5mm}
\facilities{$HST$(WFC3)}


\software{astropy \citep{2013A&A...558A..33A,2018AJ....156..123A},  
          Source Extractor \citep{1996A&AS..117..393B},
          DAOPHOT IV \citep{Stetson1987}
          }




\bibliography{biblio2}{}

\begin{thebibliography}{}
\expandafter\ifx\csname natexlab\endcsname\relax\def\natexlab#1{#1}\fi
\providecommand{\url}[1]{\href{#1}{#1}}
\providecommand{\dodoi}[1]{doi:~\href{http://doi.org/#1}{\nolinkurl{#1}}}
\providecommand{\doeprint}[1]{\href{http://ascl.net/#1}{\nolinkurl{http://ascl.net/#1}}}
\providecommand{\doarXiv}[1]{\href{https://arxiv.org/abs/#1}{\nolinkurl{https://arxiv.org/abs/#1}}}

\bibitem[{{Astropy Collaboration} {et~al.}(2013){Astropy Collaboration},
  {Robitaille}, {Tollerud}, {Greenfield}, {Droettboom}, {Bray}, {Aldcroft},
  {Davis}, {Ginsburg}, {Price-Whelan}, {Kerzendorf}, {Conley}, {Crighton},
  {Barbary}, {Muna}, {Ferguson}, {Grollier}, {Parikh}, {Nair}, {Unther},
  {Deil}, {Woillez}, {Conseil}, {Kramer}, {Turner}, {Singer}, {Fox}, {Weaver},
  {Zabalza}, {Edwards}, {Azalee Bostroem}, {Burke}, {Casey}, {Crawford},
  {Dencheva}, {Ely}, {Jenness}, {Labrie}, {Lim}, {Pierfederici}, {Pontzen},
  {Ptak}, {Refsdal}, {Servillat}, \& {Streicher}}]{2013A&A...558A..33A}
{Astropy Collaboration}, {Robitaille}, T.~P., {Tollerud}, E.~J., {et~al.} 2013,
  \aap, 558, A33, \dodoi{10.1051/0004-6361/201322068}

\bibitem[{{Astropy Collaboration} {et~al.}(2018){Astropy Collaboration},
  {Price-Whelan}, {Sip{\H{o}}cz}, {G{\"u}nther}, {Lim}, {Crawford}, {Conseil},
  {Shupe}, {Craig}, {Dencheva}, {Ginsburg}, {VanderPlas}, {Bradley},
  {P{\'e}rez-Su{\'a}rez}, {de Val-Borro}, {Aldcroft}, {Cruz}, {Robitaille},
  {Tollerud}, {Ardelean}, {Babej}, {Bach}, {Bachetti}, {Bakanov}, {Bamford},
  {Barentsen}, {Barmby}, {Baumbach}, {Berry}, {Biscani}, {Boquien}, {Bostroem},
  {Bouma}, {Brammer}, {Bray}, {Breytenbach}, {Buddelmeijer}, {Burke},
  {Calderone}, {Cano Rodr{\'\i}guez}, {Cara}, {Cardoso}, {Cheedella}, {Copin},
  {Corrales}, {Crichton}, {D'Avella}, {Deil}, {Depagne}, {Dietrich}, {Donath},
  {Droettboom}, {Earl}, {Erben}, {Fabbro}, {Ferreira}, {Finethy}, {Fox},
  {Garrison}, {Gibbons}, {Goldstein}, {Gommers}, {Greco}, {Greenfield},
  {Groener}, {Grollier}, {Hagen}, {Hirst}, {Homeier}, {Horton}, {Hosseinzadeh},
  {Hu}, {Hunkeler}, {Ivezi{\'c}}, {Jain}, {Jenness}, {Kanarek}, {Kendrew},
  {Kern}, {Kerzendorf}, {Khvalko}, {King}, {Kirkby}, {Kulkarni}, {Kumar},
  {Lee}, {Lenz}, {Littlefair}, {Ma}, {Macleod}, {Mastropietro}, {McCully},
  {Montagnac}, {Morris}, {Mueller}, {Mumford}, {Muna}, {Murphy}, {Nelson},
  {Nguyen}, {Ninan}, {N{\"o}the}, {Ogaz}, {Oh}, {Parejko}, {Parley}, {Pascual},
  {Patil}, {Patil}, {Plunkett}, {Prochaska}, {Rastogi}, {Reddy Janga},
  {Sabater}, {Sakurikar}, {Seifert}, {Sherbert}, {Sherwood-Taylor}, {Shih},
  {Sick}, {Silbiger}, {Singanamalla}, {Singer}, {Sladen}, {Sooley},
  {Sornarajah}, {Streicher}, {Teuben}, {Thomas}, {Tremblay}, {Turner},
  {Terr{\'o}n}, {van Kerkwijk}, {de la Vega}, {Watkins}, {Weaver}, {Whitmore},
  {Woillez}, {Zabalza}, \& {Astropy Contributors}}]{2018AJ....156..123A}
{Astropy Collaboration}, {Price-Whelan}, A.~M., {Sip{\H{o}}cz}, B.~M., {et~al.}
  2018, \aj, 156, 123, \dodoi{10.3847/1538-3881/aabc4f}

\bibitem[{{Barbuy} {et~al.}(2018){Barbuy}, {Chiappini}, \&
  {Gerhard}}]{barbuy18}
{Barbuy}, B., {Chiappini}, C., \& {Gerhard}, O. 2018, \araa, 56, 223,
  \dodoi{10.1146/annurev-astro-081817-051826}

\bibitem[{{Baumgardt} \& {Hilker}(2018)}]{Baumgardt2018}
{Baumgardt}, H., \& {Hilker}, M. 2018, \mnras, 478, 1520,
  \dodoi{10.1093/mnras/sty1057}

\bibitem[{{Baumgardt} {et~al.}(2019){Baumgardt}, {Hilker}, {Sollima}, \&
  {Bellini}}]{Baumgardt2019}
{Baumgardt}, H., {Hilker}, M., {Sollima}, A., \& {Bellini}, A. 2019, \mnras,
  482, 5138, \dodoi{10.1093/mnras/sty2997}

\bibitem[{{Baumgardt} \& {Vasiliev}(2021)}]{Baumgardt2021}
{Baumgardt}, H., \& {Vasiliev}, E. 2021, \mnras, 505, 5957,
  \dodoi{10.1093/mnras/stab1474}

\bibitem[{{Bellini} {et~al.}(2011){Bellini}, {Anderson}, \&
  {Bedin}}]{Bellini2011}
{Bellini}, A., {Anderson}, J., \& {Bedin}, L.~R. 2011, \pasp, 123, 622,
  \dodoi{10.1086/659878}

\bibitem[{{Bertin} \& {Arnouts}(1996)}]{1996A&AS..117..393B}
{Bertin}, E., \& {Arnouts}, S. 1996, \aaps, 117, 393,
  \dodoi{10.1051/aas:1996164}

\bibitem[{{Bhat} {et~al.}(2022){Bhat}, {Lanzoni}, {Ferraro}, \&
  {Vesperini}}]{bhat22}
{Bhat}, B., {Lanzoni}, B., {Ferraro}, F.~R., \& {Vesperini}, E. 2022, \apj,
  926, 118, \dodoi{10.3847/1538-4357/ac43bf}

\bibitem[{{Bica} {et~al.}(2006){Bica}, {Bonatto}, {Barbuy}, \&
  {Ortolani}}]{bica06}
{Bica}, E., {Bonatto}, C., {Barbuy}, B., \& {Ortolani}, S. 2006, \aap, 450,
  105, \dodoi{10.1051/0004-6361:20054351}

\bibitem[{{Brogaard} {et~al.}(2017){Brogaard}, {VandenBerg}, {Bedin}, {Milone},
  {Thygesen}, \& {Grundahl}}]{Brogaard2017}
{Brogaard}, K., {VandenBerg}, D.~A., {Bedin}, L.~R., {et~al.} 2017, \mnras,
  468, 645, \dodoi{10.1093/mnras/stx378}

\bibitem[{{Cadelano} {et~al.}(2020{\natexlab{a}}){Cadelano}, {Chen},
  {Pallanca}, {Istrate}, {Ferraro}, {Lanzoni}, {Freire}, \&
  {Salaris}}]{cadelano20psr}
{Cadelano}, M., {Chen}, J., {Pallanca}, C., {et~al.} 2020{\natexlab{a}}, \apj,
  905, 63, \dodoi{10.3847/1538-4357/abc345}

\bibitem[{{Cadelano} {et~al.}(2017){Cadelano}, {Dalessandro}, {Ferraro},
  {Miocchi}, {Lanzoni}, {Pallanca}, \& {Massari}}]{Cadelano2017}
{Cadelano}, M., {Dalessandro}, E., {Ferraro}, F.~R., {et~al.} 2017, \apj, 836,
  170, \dodoi{10.3847/1538-4357/aa5ca5}

\bibitem[{{Cadelano} {et~al.}(2019){Cadelano}, {Ferraro}, {Istrate},
  {Pallanca}, {Lanzoni}, \& {Freire}}]{Cadelano2019}
{Cadelano}, M., {Ferraro}, F.~R., {Istrate}, A.~G., {et~al.} 2019, \apj, 875,
  25, \dodoi{10.3847/1538-4357/ab0e6b}

\bibitem[{{Cadelano} {et~al.}(2018){Cadelano}, {Ransom}, {Freire}, {Ferraro},
  {Hessels}, {Lanzoni}, {Pallanca}, \& {Stairs}}]{Cadelano2018}
{Cadelano}, M., {Ransom}, S.~M., {Freire}, P.~C.~C., {et~al.} 2018, \apj, 855,
  125, \dodoi{10.3847/1538-4357/aaac2a}

\bibitem[{{Cadelano} {et~al.}(2020{\natexlab{b}}){Cadelano}, {Saracino},
  {Dalessandro}, {Ferraro}, {Lanzoni}, {Massari}, {Pallanca}, \&
  {Salaris}}]{Cadelano2020}
{Cadelano}, M., {Saracino}, S., {Dalessandro}, E., {et~al.} 2020{\natexlab{b}},
  \apj, 895, 54, \dodoi{10.3847/1538-4357/ab88b3}

\bibitem[{{Cardelli} {et~al.}(1989){Cardelli}, {Clayton}, \&
  {Mathis}}]{Cardelli1989}
{Cardelli}, J.~A., {Clayton}, G.~C., \& {Mathis}, J.~S. 1989, \apj, 345, 245,
  \dodoi{10.1086/167900}

\bibitem[{{Carretta} {et~al.}(2009){Carretta}, {Bragaglia}, {Gratton},
  {D'Orazi}, \& {Lucatello}}]{Carretta2009}
{Carretta}, E., {Bragaglia}, A., {Gratton}, R., {D'Orazi}, V., \& {Lucatello},
  S. 2009, \aap, 508, 695, \dodoi{10.1051/0004-6361/200913003}

\bibitem[{{Casagrande} \& {VandenBerg}(2014)}]{Casagrande2014}
{Casagrande}, L., \& {VandenBerg}, D.~A. 2014, \mnras, 444, 392,
  \dodoi{10.1093/mnras/stu1476}

\bibitem[{{Chun} {et~al.}(2015){Chun}, {Kang}, {Jung}, \& {Sohn}}]{Chun2015}
{Chun}, S.-H., {Kang}, M., {Jung}, D., \& {Sohn}, Y.-J. 2015, \aj, 149, 29,
  \dodoi{10.1088/0004-6256/149/1/29}

\bibitem[{{Clarkson} {et~al.}(2011){Clarkson}, {Sahu}, {Anderson}, {Rich},
  {Smith}, {Brown}, {Bond}, {Livio}, {Minniti}, {Renzini}, \&
  {Zoccali}}]{Clarkson2011}
{Clarkson}, W.~I., {Sahu}, K.~C., {Anderson}, J., {et~al.} 2011, \apj, 735, 37,
  \dodoi{10.1088/0004-637X/735/1/37}

\bibitem[{{Cohen} {et~al.}(2021){Cohen}, {Bellini}, {Casagrande}, {Brown},
  {Correnti}, \& {Kalirai}}]{Cohen2021}
{Cohen}, R.~E., {Bellini}, A., {Casagrande}, L., {et~al.} 2021, \aj, 162, 228,
  \dodoi{10.3847/1538-3881/ac281f}

\bibitem[{{Conroy} {et~al.}(2018){Conroy}, {Villaume}, {van Dokkum}, \&
  {Lind}}]{Conroy2018}
{Conroy}, C., {Villaume}, A., {van Dokkum}, P.~G., \& {Lind}, K. 2018, \apj,
  854, 139, \dodoi{10.3847/1538-4357/aaab49}

\bibitem[{{Davidge} {et~al.}(1992){Davidge}, {Harris}, {Bridges}, \&
  {Hanes}}]{Davidge1992}
{Davidge}, T.~J., {Harris}, W.~E., {Bridges}, T.~J., \& {Hanes}, D.~A. 1992,
  \apjs, 81, 251, \dodoi{10.1086/191692}

\bibitem[{{Dias} {et~al.}(2016){Dias}, {Barbuy}, {Saviane}, {Held}, {Da Costa},
  {Ortolani}, {Gullieuszik}, \& {V{\'a}squez}}]{Dias2016}
{Dias}, B., {Barbuy}, B., {Saviane}, I., {et~al.} 2016, \aap, 590, A9,
  \dodoi{10.1051/0004-6361/201526765}

\bibitem[{{Djorgovski}(1993)}]{Djorgovski1993}
{Djorgovski}, S. 1993, in Astronomical Society of the Pacific Conference
  Series, Vol.~50, Structure and Dynamics of Globular Clusters, ed. S.~G.
  {Djorgovski} \& G.~{Meylan}, 373

\bibitem[{{Dotter} {et~al.}(2008){Dotter}, {Chaboyer}, {Jevremovi{\'c}},
  {Kostov}, {Baron}, \& {Ferguson}}]{Dotter2008}
{Dotter}, A., {Chaboyer}, B., {Jevremovi{\'c}}, D., {et~al.} 2008, \apjs, 178,
  89, \dodoi{10.1086/589654}

\bibitem[{{Dotter} {et~al.}(2010){Dotter}, {Sarajedini}, {Anderson},
  {Aparicio}, {Bedin}, {Chaboyer}, {Majewski}, {Mar{\'\i}n-Franch}, {Milone},
  {Paust}, {Piotto}, {Reid}, {Rosenberg}, \& {Siegel}}]{Dotter2010}
{Dotter}, A., {Sarajedini}, A., {Anderson}, J., {et~al.} 2010, \apj, 708, 698,
  \dodoi{10.1088/0004-637X/708/1/698}

\bibitem[{{Ferraro}(2017)}]{Ferraro2017}
{Ferraro}, F. 2017, {Pushing ahead the frontier of the Globular Cluster
  dynamics: the 3D view of the velocity space}, HST Proposal

\bibitem[{{Ferraro} {et~al.}(2020){Ferraro}, {Lanzoni}, \&
  {Dalessandro}}]{Ferraro2020}
{Ferraro}, F.~R., {Lanzoni}, B., \& {Dalessandro}, E. 2020, Rendiconti Lincei.
  Scienze Fisiche e Naturali, 31, 19, \dodoi{10.1007/s12210-020-00873-2}

\bibitem[{{Ferraro} {et~al.}(2019){Ferraro}, {Lanzoni}, {Dalessandro},
  {Cadelano}, {Raso}, {Mucciarelli}, {Beccari}, \& {Pallanca}}]{Ferraro2019}
{Ferraro}, F.~R., {Lanzoni}, B., {Dalessandro}, E., {et~al.} 2019, Nature
  Astronomy, 3, 1149, \dodoi{10.1038/s41550-019-0865-1}

\bibitem[{{Ferraro} {et~al.}(2016){Ferraro}, {Massari}, {Dalessandro},
  {Lanzoni}, {Origlia}, {Rich}, \& {Mucciarelli}}]{Ferraro2016}
{Ferraro}, F.~R., {Massari}, D., {Dalessandro}, E., {et~al.} 2016, \apj, 828,
  75, \dodoi{10.3847/0004-637X/828/2/75}

\bibitem[{{Ferraro} {et~al.}(2009){Ferraro}, {Dalessandro}, {Mucciarelli},
  {Beccari}, {Rich}, {Origlia}, {Lanzoni}, {Rood}, {Valenti}, {Bellazzini},
  {Ransom}, \& {Cocozza}}]{Ferraro2009}
{Ferraro}, F.~R., {Dalessandro}, E., {Mucciarelli}, A., {et~al.} 2009, \nat,
  462, 483, \dodoi{10.1038/nature08581}

\bibitem[{{Ferraro} {et~al.}(2012){Ferraro}, {Lanzoni}, {Dalessandro},
  {Beccari}, {Pasquato}, {Miocchi}, {Rood}, {Sigurdsson}, {Sills}, {Vesperini},
  {Mapelli}, {Contreras}, {Sanna}, \& {Mucciarelli}}]{Ferraro2012}
{Ferraro}, F.~R., {Lanzoni}, B., {Dalessandro}, E., {et~al.} 2012, \nat, 492,
  393, \dodoi{10.1038/nature11686}

\bibitem[{{Ferraro} {et~al.}(2018){Ferraro}, {Lanzoni}, {Raso}, {Nardiello},
  {Dalessandro}, {Vesperini}, {Piotto}, {Pallanca}, {Beccari}, {Bellini},
  {Libralato}, {Anderson}, {Aparicio}, {Bedin}, {Cassisi}, {Milone},
  {Ortolani}, {Renzini}, {Salaris}, \& {van der Marel}}]{Ferraro2018}
{Ferraro}, F.~R., {Lanzoni}, B., {Raso}, S., {et~al.} 2018, \apj, 860, 36,
  \dodoi{10.3847/1538-4357/aac01c}

\bibitem[{{Ferraro} {et~al.}(2021){Ferraro}, {Pallanca}, {Lanzoni}, {Crociati},
  {Dalessandro}, {Origlia}, {Rich}, {Saracino}, {Mucciarelli}, {Valenti},
  {Geisler}, {Mauro}, {Villanova}, {Moni Bidin}, \& {Beccari}}]{Ferraro2021}
{Ferraro}, F.~R., {Pallanca}, C., {Lanzoni}, B., {et~al.} 2021, Nature
  Astronomy, 5, 311, \dodoi{10.1038/s41550-020-01267-y}

\bibitem[{{Forbes} {et~al.}(2018){Forbes}, {Bastian}, {Gieles}, {Crain},
  {Kruijssen}, {Larsen}, {Ploeckinger}, {Agertz}, {Trenti}, {Ferguson},
  {Pfeffer}, \& {Gnedin}}]{Forbes2018}
{Forbes}, D.~A., {Bastian}, N., {Gieles}, M., {et~al.} 2018, Proceedings of the
  Royal Society of London Series A, 474, 20170616,
  \dodoi{10.1098/rspa.2017.0616}

\bibitem[{{Foreman-Mackey} {et~al.}(2013){Foreman-Mackey}, {Hogg}, {Lang}, \&
  {Goodman}}]{foreman13}
{Foreman-Mackey}, D., {Hogg}, D.~W., {Lang}, D., \& {Goodman}, J. 2013, \pasp,
  125, 306, \dodoi{10.1086/670067}

\bibitem[{{Foreman-Mackey} {et~al.}(2019){Foreman-Mackey}, {Farr}, {Sinha},
  {Archibald}, {Hogg}, {Sanders}, {Zuntz}, {Williams}, {Nelson}, {de
  Val-Borro}, {Erhardt}, {Pashchenko}, \& {Pla}}]{foreman19}
{Foreman-Mackey}, D., {Farr}, W., {Sinha}, M., {et~al.} 2019, The Journal of
  Open Source Software, 4, 1864, \dodoi{10.21105/joss.01864}

\bibitem[{{Gaia Collaboration} {et~al.}(2016){Gaia Collaboration}, {Prusti},
  {de Bruijne}, {Brown}, {Vallenari}, {Babusiaux}, {Bailer-Jones}, {Bastian},
  {Biermann}, {Evans}, {Eyer}, {Jansen}, {Jordi}, {Klioner}, {Lammers},
  {Lindegren}, {Luri}, {Mignard}, {Milligan}, {Panem}, {Poinsignon},
  {Pourbaix}, {Randich}, {Sarri}, {Sartoretti}, {Siddiqui}, {Soubiran},
  {Valette}, {van Leeuwen}, {Walton}, {Aerts}, {Arenou}, {Cropper}, {Drimmel},
  {H{\o}g}, {Katz}, {Lattanzi}, {O'Mullane}, {Grebel}, {Holland}, {Huc},
  {Passot}, {Bramante}, {Cacciari}, {Casta{\~n}eda}, {Chaoul}, {Cheek}, {De
  Angeli}, {Fabricius}, {Guerra}, {Hern{\'a}ndez}, {Jean-Antoine-Piccolo},
  {Masana}, {Messineo}, {Mowlavi}, {Nienartowicz}, {Ord{\'o}{\~n}ez-Blanco},
  {Panuzzo}, {Portell}, {Richards}, {Riello}, {Seabroke}, {Tanga},
  {Th{\'e}venin}, {Torra}, {Els}, {Gracia-Abril}, {Comoretto},
  {Garcia-Reinaldos}, {Lock}, {Mercier}, {Altmann}, {Andrae}, {Astraatmadja},
  {Bellas-Velidis}, {Benson}, {Berthier}, {Blomme}, {Busso}, {Carry},
  {Cellino}, {Clementini}, {Cowell}, {Creevey}, {Cuypers}, {Davidson}, {De
  Ridder}, {de Torres}, {Delchambre}, {Dell'Oro}, {Ducourant}, {Fr{\'e}mat},
  {Garc{\'\i}a-Torres}, {Gosset}, {Halbwachs}, {Hambly}, {Harrison}, {Hauser},
  {Hestroffer}, {Hodgkin}, {Huckle}, {Hutton}, {Jasniewicz}, {Jordan},
  {Kontizas}, {Korn}, {Lanzafame}, {Manteiga}, {Moitinho}, {Muinonen},
  {Osinde}, {Pancino}, {Pauwels}, {Petit}, {Recio-Blanco}, {Robin}, {Sarro},
  {Siopis}, {Smith}, {Smith}, {Sozzetti}, {Thuillot}, {van Reeven}, {Viala},
  {Abbas}, {Abreu Aramburu}, {Accart}, {Aguado}, {Allan}, {Allasia},
  {Altavilla}, {{\'A}lvarez}, {Alves}, {Anderson}, {Andrei}, {Anglada Varela},
  {Antiche}, {Antoja}, {Ant{\'o}n}, {Arcay}, {Atzei}, {Ayache}, {Bach},
  {Baker}, {Balaguer-N{\'u}{\~n}ez}, {Barache}, {Barata}, {Barbier}, {Barblan},
  {Baroni}, {Barrado y Navascu{\'e}s}, {Barros}, {Barstow}, {Becciani},
  {Bellazzini}, {Bellei}, {Bello Garc{\'\i}a}, {Belokurov}, {Bendjoya},
  {Berihuete}, {Bianchi}, {Bienaym{\'e}}, {Billebaud}, {Blagorodnova},
  {Blanco-Cuaresma}, {Boch}, {Bombrun}, {Borrachero}, {Bouquillon}, {Bourda},
  {Bouy}, {Bragaglia}, {Breddels}, {Brouillet}, {Br{\"u}semeister},
  {Bucciarelli}, {Budnik}, {Burgess}, {Burgon}, {Burlacu}, {Busonero}, {Buzzi},
  {Caffau}, {Cambras}, {Campbell}, {Cancelliere}, {Cantat-Gaudin}, {Carlucci},
  {Carrasco}, {Castellani}, {Charlot}, {Charnas}, {Charvet}, {Chassat},
  {Chiavassa}, {Clotet}, {Cocozza}, {Collins}, {Collins}, {Costigan}, {Crifo},
  {Cross}, {Crosta}, {Crowley}, {Dafonte}, {Damerdji}, {Dapergolas}, {David},
  {David}, {De Cat}, {de Felice}, {de Laverny}, {De Luise}, {De March}, {de
  Martino}, {de Souza}, {Debosscher}, {del Pozo}, {Delbo}, {Delgado},
  {Delgado}, {di Marco}, {Di Matteo}, {Diakite}, {Distefano}, {Dolding}, {Dos
  Anjos}, {Drazinos}, {Dur{\'a}n}, {Dzigan}, {Ecale}, {Edvardsson}, {Enke},
  {Erdmann}, {Escolar}, {Espina}, {Evans}, {Eynard Bontemps}, {Fabre},
  {Fabrizio}, {Faigler}, {Falc{\~a}o}, {Farr{\`a}s Casas}, {Faye}, {Federici},
  {Fedorets}, {Fern{\'a}ndez-Hern{\'a}ndez}, {Fernique}, {Fienga}, {Figueras},
  {Filippi}, {Findeisen}, {Fonti}, {Fouesneau}, {Fraile}, {Fraser}, {Fuchs},
  {Furnell}, {Gai}, {Galleti}, {Galluccio}, {Garabato}, {Garc{\'\i}a-Sedano},
  {Gar{\'e}}, {Garofalo}, {Garralda}, {Gavras}, {Gerssen}, {Geyer}, {Gilmore},
  {Girona}, {Giuffrida}, {Gomes}, {Gonz{\'a}lez-Marcos},
  {Gonz{\'a}lez-N{\'u}{\~n}ez}, {Gonz{\'a}lez-Vidal}, {Granvik}, {Guerrier},
  {Guillout}, {Guiraud}, {G{\'u}rpide}, {Guti{\'e}rrez-S{\'a}nchez}, {Guy},
  {Haigron}, {Hatzidimitriou}, {Haywood}, {Heiter}, {Helmi}, {Hobbs},
  {Hofmann}, {Holl}, {Holland}, {Hunt}, {Hypki}, {Icardi}, {Irwin}, {Jevardat
  de Fombelle}, {Jofr{\'e}}, {Jonker}, {Jorissen}, {Julbe}, {Karampelas},
  {Kochoska}, {Kohley}, {Kolenberg}, {Kontizas}, {Koposov}, {Kordopatis},
  {Koubsky}, {Kowalczyk}, {Krone-Martins}, {Kudryashova}, {Kull}, {Bachchan},
  {Lacoste-Seris}, {Lanza}, {Lavigne}, {Le Poncin-Lafitte}, {Lebreton},
  {Lebzelter}, {Leccia}, {Leclerc}, {Lecoeur-Taibi}, {Lemaitre}, {Lenhardt},
  {Leroux}, {Liao}, {Licata}, {Lindstr{\o}m}, {Lister}, {Livanou}, {Lobel},
  {L{\"o}ffler}, {L{\'o}pez}, {Lopez-Lozano}, {Lorenz}, {Loureiro},
  {MacDonald}, {Magalh{\~a}es Fernandes}, {Managau}, {Mann}, {Mantelet},
  {Marchal}, {Marchant}, {Marconi}, {Marie}, {Marinoni}, {Marrese},
  {Marschalk{\'o}}, {Marshall}, {Mart{\'\i}n-Fleitas}, {Martino}, {Mary},
  {Matijevi{\v{c}}}, {Mazeh}, {McMillan}, {Messina}, {Mestre}, {Michalik},
  {Millar}, {Miranda}, {Molina}, {Molinaro}, {Molinaro}, {Moln{\'a}r},
  {Moniez}, {Montegriffo}, {Monteiro}, {Mor}, {Mora}, {Morbidelli}, {Morel},
  {Morgenthaler}, {Morley}, {Morris}, {Mulone}, {Muraveva}, {Musella},
  {Narbonne}, {Nelemans}, {Nicastro}, {Noval}, {Ord{\'e}novic},
  {Ordieres-Mer{\'e}}, {Osborne}, {Pagani}, {Pagano}, {Pailler}, {Palacin},
  {Palaversa}, {Parsons}, {Paulsen}, {Pecoraro}, {Pedrosa}, {Pentik{\"a}inen},
  {Pereira}, {Pichon}, {Piersimoni}, {Pineau}, {Plachy}, {Plum}, {Poujoulet},
  {Pr{\v{s}}a}, {Pulone}, {Ragaini}, {Rago}, {Rambaux}, {Ramos-Lerate},
  {Ranalli}, {Rauw}, {Read}, {Regibo}, {Renk}, {Reyl{\'e}}, {Ribeiro},
  {Rimoldini}, {Ripepi}, {Riva}, {Rixon}, {Roelens}, {Romero-G{\'o}mez},
  {Rowell}, {Royer}, {Rudolph}, {Ruiz-Dern}, {Sadowski}, {Sagrist{\`a}
  Sell{\'e}s}, {Sahlmann}, {Salgado}, {Salguero}, {Sarasso}, {Savietto},
  {Schnorhk}, {Schultheis}, {Sciacca}, {Segol}, {Segovia}, {Segransan},
  {Serpell}, {Shih}, {Smareglia}, {Smart}, {Smith}, {Solano}, {Solitro},
  {Sordo}, {Soria Nieto}, {Souchay}, {Spagna}, {Spoto}, {Stampa}, {Steele},
  {Steidelm{\"u}ller}, {Stephenson}, {Stoev}, {Suess}, {S{\"u}veges}, {Surdej},
  {Szabados}, {Szegedi-Elek}, {Tapiador}, {Taris}, {Tauran}, {Taylor},
  {Teixeira}, {Terrett}, {Tingley}, {Trager}, {Turon}, {Ulla}, {Utrilla},
  {Valentini}, {van Elteren}, {Van Hemelryck}, {van Leeuwen}, {Varadi},
  {Vecchiato}, {Veljanoski}, {Via}, {Vicente}, {Vogt}, {Voss}, {Votruba},
  {Voutsinas}, {Walmsley}, {Weiler}, {Weingrill}, {Werner}, {Wevers},
  {Whitehead}, {Wyrzykowski}, {Yoldas}, {{\v{Z}}erjal}, {Zucker}, {Zurbach},
  {Zwitter}, {Alecu}, {Allen}, {Allende Prieto}, {Amorim},
  {Anglada-Escud{\'e}}, {Arsenijevic}, {Azaz}, {Balm}, {Beck}, {Bernstein},
  {Bigot}, {Bijaoui}, {Blasco}, {Bonfigli}, {Bono}, {Boudreault}, {Bressan},
  {Brown}, {Brunet}, {Bunclark}, {Buonanno}, {Butkevich}, {Carret}, {Carrion},
  {Chemin}, {Ch{\'e}reau}, {Corcione}, {Darmigny}, {de Boer}, {de Teodoro}, {de
  Zeeuw}, {Delle Luche}, {Domingues}, {Dubath}, {Fodor}, {Fr{\'e}zouls},
  {Fries}, {Fustes}, {Fyfe}, {Gallardo}, {Gallegos}, {Gardiol}, {Gebran},
  {Gomboc}, {G{\'o}mez}, {Grux}, {Gueguen}, {Heyrovsky}, {Hoar}, {Iannicola},
  {Isasi Parache}, {Janotto}, {Joliet}, {Jonckheere}, {Keil}, {Kim},
  {Klagyivik}, {Klar}, {Knude}, {Kochukhov}, {Kolka}, {Kos}, {Kutka}, {Lainey},
  {LeBouquin}, {Liu}, {Loreggia}, {Makarov}, {Marseille}, {Martayan},
  {Martinez-Rubi}, {Massart}, {Meynadier}, {Mignot}, {Munari}, {Nguyen},
  {Nordlander}, {Ocvirk}, {O'Flaherty}, {Olias Sanz}, {Ortiz}, {Osorio},
  {Oszkiewicz}, {Ouzounis}, {Palmer}, {Park}, {Pasquato}, {Peltzer}, {Peralta},
  {P{\'e}turaud}, {Pieniluoma}, {Pigozzi}, {Poels}, {Prat}, {Prod'homme},
  {Raison}, {Rebordao}, {Risquez}, {Rocca-Volmerange}, {Rosen}, {Ruiz-Fuertes},
  {Russo}, {Sembay}, {Serraller Vizcaino}, {Short}, {Siebert}, {Silva},
  {Sinachopoulos}, {Slezak}, {Soffel}, {Sosnowska}, {Strai{\v{z}}ys}, {ter
  Linden}, {Terrell}, {Theil}, {Tiede}, {Troisi}, {Tsalmantza}, {Tur},
  {Vaccari}, {Vachier}, {Valles}, {Van Hamme}, {Veltz}, {Virtanen}, {Wallut},
  {Wichmann}, {Wilkinson}, {Ziaeepour}, \& {Zschocke}}]{Gaia2016}
{Gaia Collaboration}, {Prusti}, T., {de Bruijne}, J.~H.~J., {et~al.} 2016,
  \aap, 595, A1, \dodoi{10.1051/0004-6361/201629272}

\bibitem[{{Gaia Collaboration} {et~al.}(2022){Gaia Collaboration}, {Vallenari},
  {Brown}, {Prusti}, {de Bruijne}, {Arenou}, {Babusiaux}, {Biermann},
  {Creevey}, {Ducourant}, {Evans}, {Eyer}, {Guerra}, {Hutton}, {Jordi},
  {Klioner}, {Lammers}, {Lindegren}, {Luri}, {Mignard}, {Panem}, {Pourbaix},
  {Randich}, {Sartoretti}, {Soubiran}, {Tanga}, {Walton}, {Bailer-Jones},
  {Bastian}, {Drimmel}, {Jansen}, {Katz}, {Lattanzi}, {van Leeuwen}, {Bakker},
  {Cacciari}, {Casta{\~n}eda}, {De Angeli}, {Fabricius}, {Fouesneau},
  {Fr{\'e}mat}, {Galluccio}, {Guerrier}, {Heiter}, {Masana}, {Messineo},
  {Mowlavi}, {Nicolas}, {Nienartowicz}, {Pailler}, {Panuzzo}, {Riclet}, {Roux},
  {Seabroke}, {Sordo{\o}rcit}, {Th{\'e}venin}, {Gracia-Abril}, {Portell},
  {Teyssier}, {Altmann}, {Andrae}, {Audard}, {Bellas-Velidis}, {Benson},
  {Berthier}, {Blomme}, {Burgess}, {Busonero}, {Busso}, {C{\'a}novas}, {Carry},
  {Cellino}, {Cheek}, {Clementini}, {Damerdji}, {Davidson}, {de Teodoro},
  {Nu{\~n}ez Campos}, {Delchambre}, {Dell'Oro}, {Esquej},
  {Fern{\'a}ndez-Hern{\'a}ndez}, {Fraile}, {Garabato}, {Garc{\'\i}a-Lario},
  {Gosset}, {Haigron}, {Halbwachs}, {Hambly}, {Harrison}, {Hern{\'a}ndez},
  {Hestroffer}, {Hodgkin}, {Holl}, {Jan{\ss}en}, {Jevardat de Fombelle},
  {Jordan}, {Krone-Martins}, {Lanzafame}, {L{\"o}ffler}, {Marchal}, {Marrese},
  {Moitinho}, {Muinonen}, {Osborne}, {Pancino}, {Pauwels}, {Recio-Blanco},
  {Reyl{\'e}}, {Riello}, {Rimoldini}, {Roegiers}, {Rybizki}, {Sarro}, {Siopis},
  {Smith}, {Sozzetti}, {Utrilla}, {van Leeuwen}, {Abbas}, {{\'A}brah{\'a}m},
  {Abreu Aramburu}, {Aerts}, {Aguado}, {Ajaj}, {Aldea-Montero}, {Altavilla},
  {{\'A}lvarez}, {Alves}, {Anders}, {Anderson}, {Anglada Varela}, {Antoja},
  {Baines}, {Baker}, {Balaguer-N{\'u}{\~n}ez}, {Balbinot}, {Balog}, {Barache},
  {Barbato}, {Barros}, {Barstow}, {Bartolom{\'e}}, {Bassilana}, {Bauchet},
  {Becciani}, {Bellazzini}, {Berihuete}, {Bernet}, {Bertone}, {Bianchi},
  {Binnenfeld}, {Blanco-Cuaresma}, {Blazere}, {Boch}, {Bombrun}, {Bossini},
  {Bouquillon}, {Bragaglia}, {Bramante}, {Breedt}, {Bressan}, {Brouillet},
  {Brugaletta}, {Bucciarelli}, {Burlacu}, {Butkevich}, {Buzzi}, {Caffau},
  {Cancelliere}, {Cantat-Gaudin}, {Carballo}, {Carlucci}, {Carnerero},
  {Carrasco}, {Casamiquela}, {Castellani}, {Castro-Ginard}, {Chaoul},
  {Charlot}, {Chemin}, {Chiaramida}, {Chiavassa}, {Chornay}, {Comoretto},
  {Contursi}, {Cooper}, {Cornez}, {Cowell}, {Crifo}, {Cropper}, {Crosta},
  {Crowley}, {Dafonte}, {Dapergolas}, {David}, {David}, {de Laverny}, {De
  Luise}, {De March}, {De Ridder}, {de Souza}, {de Torres}, {del Peloso}, {del
  Pozo}, {Delbo}, {Delgado}, {Delisle}, {Demouchy}, {Dharmawardena}, {Di
  Matteo}, {Diakite}, {Diener}, {Distefano}, {Dolding}, {Edvardsson}, {Enke},
  {Fabre}, {Fabrizio}, {Faigler}, {Fedorets}, {Fernique}, {Fienga}, {Figueras},
  {Fournier}, {Fouron}, {Fragkoudi}, {Gai}, {Garcia-Gutierrez},
  {Garcia-Reinaldos}, {Garc{\'\i}a-Torres}, {Garofalo}, {Gavel}, {Gavras},
  {Gerlach}, {Geyer}, {Giacobbe}, {Gilmore}, {Girona}, {Giuffrida}, {Gomel},
  {Gomez}, {Gonz{\'a}lez-N{\'u}{\~n}ez}, {Gonz{\'a}lez-Santamar{\'\i}a},
  {Gonz{\'a}lez-Vidal}, {Granvik}, {Guillout}, {Guiraud},
  {Guti{\'e}rrez-S{\'a}nchez}, {Guy}, {Hatzidimitriou}, {Hauser}, {Haywood},
  {Helmer}, {Helmi}, {Sarmiento}, {Hidalgo}, {Hilger}, {H{\l}adczuk}, {Hobbs},
  {Holland}, {Huckle}, {Jardine}, {Jasniewicz}, {Jean-Antoine Piccolo},
  {Jim{\'e}nez-Arranz}, {Jorissen}, {Juaristi Campillo}, {Julbe}, {Karbevska},
  {Kervella}, {Khanna}, {Kontizas}, {Kordopatis}, {Korn}, {K{\'o}sp{\'a}l},
  {Kostrzewa-Rutkowska}, {Kruszy{\'n}ska}, {Kun}, {Laizeau}, {Lambert},
  {Lanza}, {Lasne}, {Le Campion}, {Lebreton}, {Lebzelter}, {Leccia}, {Leclerc},
  {Lecoeur-Taibi}, {Liao}, {Licata}, {Lindstr{\o}m}, {Lister}, {Livanou},
  {Lobel}, {Lorca}, {Loup}, {Madrero Pardo}, {Magdaleno Romeo}, {Managau},
  {Mann}, {Manteiga}, {Marchant}, {Marconi}, {Marcos}, {Marcos Santos},
  {Mar{\'\i}n Pina}, {Marinoni}, {Marocco}, {Marshall}, {Polo},
  {Mart{\'\i}n-Fleitas}, {Marton}, {Mary}, {Masip}, {Massari},
  {Mastrobuono-Battisti}, {Mazeh}, {McMillan}, {Messina}, {Michalik}, {Millar},
  {Mints}, {Molina}, {Molinaro}, {Moln{\'a}r}, {Monari}, {Mongui{\'o}},
  {Montegriffo}, {Montero}, {Mor}, {Mora}, {Morbidelli}, {Morel}, {Morris},
  {Muraveva}, {Murphy}, {Musella}, {Nagy}, {Noval}, {Oca{\~n}a}, {Ogden},
  {Ordenovic}, {Osinde}, {Pagani}, {Pagano}, {Palaversa}, {Palicio},
  {Pallas-Quintela}, {Panahi}, {Payne-Wardenaar}, {Pe{\~n}alosa Esteller},
  {Penttil{\"a}}, {Pichon}, {Piersimoni}, {Pineau}, {Plachy}, {Plum}, {Poggio},
  {Pr{\v{s}}a}, {Pulone}, {Racero}, {Ragaini}, {Rainer}, {Raiteri}, {Rambaux},
  {Ramos}, {Ramos-Lerate}, {Re Fiorentin}, {Regibo}, {Richards}, {Rios Diaz},
  {Ripepi}, {Riva}, {Rix}, {Rixon}, {Robichon}, {Robin}, {Robin}, {Roelens},
  {Rogues}, {Rohrbasser}, {Romero-G{\'o}mez}, {Rowell}, {Royer}, {Ruz Mieres},
  {Rybicki}, {Sadowski}, {S{\'a}ez N{\'u}{\~n}ez}, {Sagrist{\`a} Sell{\'e}s},
  {Sahlmann}, {Salguero}, {Samaras}, {Sanchez Gimenez}, {Sanna},
  {Santove{\~n}a}, {Sarasso}, {Schultheis}, {Sciacca}, {Segol}, {Segovia},
  {S{\'e}gransan}, {Semeux}, {Shahaf}, {Siddiqui}, {Siebert}, {Siltala},
  {Silvelo}, {Slezak}, {Slezak}, {Smart}, {Snaith}, {Solano}, {Solitro},
  {Souami}, {Souchay}, {Spagna}, {Spina}, {Spoto}, {Steele},
  {Steidelm{\"u}ller}, {Stephenson}, {S{\"u}veges}, {Surdej}, {Szabados},
  {Szegedi-Elek}, {Taris}, {Taylo}, {Teixeira}, {Tolomei}, {Tonello}, {Torra},
  {Torra}, {Torralba Elipe}, {Trabucchi}, {Tsounis}, {Turon}, {Ulla}, {Unger},
  {Vaillant}, {van Dillen}, {van Reeven}, {Vanel}, {Vecchiato}, {Viala},
  {Vicente}, {Voutsinas}, {Weiler}, {Wevers}, {Wyrzykowski}, {Yoldas}, {Yvard},
  {Zhao}, {Zorec}, {Zucker}, \& {Zwitter}}]{Gaia2022}
{Gaia Collaboration}, {Vallenari}, A., {Brown}, A.~G.~A., {et~al.} 2022, arXiv
  e-prints, arXiv:2208.00211.
\newblock \doarXiv{2208.00211}

\bibitem[{{Girardi} {et~al.}(2002){Girardi}, {Bertelli}, {Bressan}, {Chiosi},
  {Groenewegen}, {Marigo}, {Salasnich}, \& {Weiss}}]{Girardi2002}
{Girardi}, L., {Bertelli}, G., {Bressan}, A., {et~al.} 2002, \aap, 391, 195,
  \dodoi{10.1051/0004-6361:20020612}

\bibitem[{{Harris}(1996)}]{Harris1996}
{Harris}, W.~E. 1996, \aj, 112, 1487

\bibitem[{{Heitsch} \& {Richtler}(1999)}]{Heitsch1999}
{Heitsch}, F., \& {Richtler}, T. 1999, \aap, 347, 455.
\newblock \doarXiv{astro-ph/9904404}

\bibitem[{{Kerber} {et~al.}(2018){Kerber}, {Nardiello}, {Ortolani}, {Barbuy},
  {Bica}, {Cassisi}, {Libralato}, \& {Vieira}}]{Kerber2018}
{Kerber}, L.~O., {Nardiello}, D., {Ortolani}, S., {et~al.} 2018, \apj, 853, 15,
  \dodoi{10.3847/1538-4357/aaa3fc}

\bibitem[{{Kerber} {et~al.}(2019){Kerber}, {Libralato}, {Souza}, {Oliveira},
  {Ortolani}, {P{\'e}rez-Villegas}, {Barbuy}, {Dias}, {Bica}, \&
  {Nardiello}}]{Kerber2019}
{Kerber}, L.~O., {Libralato}, M., {Souza}, S.~O., {et~al.} 2019, \mnras, 484,
  5530, \dodoi{10.1093/mnras/stz003}

\bibitem[{{King}(1966)}]{King1966}
{King}, I.~R. 1966, \aj, 71, 64, \dodoi{10.1086/109857}

\bibitem[{{Lanzoni} {et~al.}(2007){Lanzoni}, {Dalessandro}, {Ferraro},
  {Miocchi}, {Valenti}, \& {Rood}}]{Lanzoni2007}
{Lanzoni}, B., {Dalessandro}, E., {Ferraro}, F.~R., {et~al.} 2007, \apjl, 668,
  L139, \dodoi{10.1086/522927}

\bibitem[{{Lanzoni} {et~al.}(2016){Lanzoni}, {Ferraro}, {Alessandrini},
  {Dalessandro}, {Vesperini}, \& {Raso}}]{Lanzoni2016}
{Lanzoni}, B., {Ferraro}, F.~R., {Alessandrini}, E., {et~al.} 2016, \apjl, 833,
  L29, \dodoi{10.3847/2041-8213/833/2/L29}

\bibitem[{{Lanzoni} {et~al.}(2010){Lanzoni}, {Ferraro}, {Dalessandro},
  {Mucciarelli}, {Beccari}, {Miocchi}, {Bellazzini}, {Rich}, {Origlia},
  {Valenti}, {Rood}, \& {Ransom}}]{Lanzoni2010}
{Lanzoni}, B., {Ferraro}, F.~R., {Dalessandro}, E., {et~al.} 2010, \apj, 717,
  653, \dodoi{10.1088/0004-637X/717/2/653}

\bibitem[{{Lanzoni} {et~al.}(2019){Lanzoni}, {Ferraro}, {Dalessandro},
  {Cadelano}, {Pallanca}, {Raso}, {Mucciarelli}, {Beccari}, \&
  {Focardi}}]{Lanzoni2019}
---. 2019, \apj, 887, 176, \dodoi{10.3847/1538-4357/ab54c2}

\bibitem[{{Lee} {et~al.}(2018){Lee}, {Hong}, {Lim}, {Chung}, {Jang}, {Kim}, \&
  {Joo}}]{Lee2018}
{Lee}, Y.-W., {Hong}, S., {Lim}, D., {et~al.} 2018, \apjl, 862, L8,
  \dodoi{10.3847/2041-8213/aad192}

\bibitem[{{Marigo} {et~al.}(2017){Marigo}, {Girardi}, {Bressan}, {Rosenfield},
  {Aringer}, {Chen}, {Dussin}, {Nanni}, {Pastorelli}, {Rodrigues}, {Trabucchi},
  {Bladh}, {Dalcanton}, {Groenewegen}, {Montalb{\'a}n}, \& {Wood}}]{Marigo2017}
{Marigo}, P., {Girardi}, L., {Bressan}, A., {et~al.} 2017, \apj, 835, 77,
  \dodoi{10.3847/1538-4357/835/1/77}

\bibitem[{{Massari} {et~al.}(2019){Massari}, {Koppelman}, \&
  {Helmi}}]{Massari2019}
{Massari}, D., {Koppelman}, H.~H., \& {Helmi}, A. 2019, \aap, 630, L4,
  \dodoi{10.1051/0004-6361/201936135}

\bibitem[{{Massari} {et~al.}(2014){Massari}, {Mucciarelli}, {Ferraro},
  {Origlia}, {Rich}, {Lanzoni}, {Dalessandro}, {Valenti}, {Ibata}, {Lovisi},
  {Bellazzini}, \& {Reitzel}}]{Massari2014}
{Massari}, D., {Mucciarelli}, A., {Ferraro}, F.~R., {et~al.} 2014, \apj, 795,
  22, \dodoi{10.1088/0004-637X/795/1/22}

\bibitem[{{Miocchi} {et~al.}(2013){Miocchi}, {Lanzoni}, {Ferraro},
  {Dalessandro}, {Vesperini}, {Pasquato}, {Beccari}, {Pallanca}, \&
  {Sanna}}]{Miocchi2013}
{Miocchi}, P., {Lanzoni}, B., {Ferraro}, F.~R., {et~al.} 2013, \apj, 774, 151,
  \dodoi{10.1088/0004-637X/774/2/151}

\bibitem[{{Moffat}(1969)}]{Moffat1969}
{Moffat}, A.~F.~J. 1969, \aap, 3, 455

\bibitem[{{Montegriffo} {et~al.}(1995){Montegriffo}, {Ferraro}, {Fusi Pecci},
  \& {Origlia}}]{Montegriffo1995}
{Montegriffo}, P., {Ferraro}, F.~R., {Fusi Pecci}, F., \& {Origlia}, L. 1995,
  \mnras, 276, 739, \dodoi{10.1093/mnras/276.3.739}

\bibitem[{{Nataf} {et~al.}(2013){Nataf}, {Gould}, {Fouqu{\'e}}, {Gonzalez},
  {Johnson}, {Skowron}, {Udalski}, {Szyma{\'n}ski}, {Kubiak},
  {Pietrzy{\'n}ski}, {Soszy{\'n}ski}, {Ulaczyk}, {Wyrzykowski}, \&
  {Poleski}}]{Nataf2013}
{Nataf}, D.~M., {Gould}, A., {Fouqu{\'e}}, P., {et~al.} 2013, \apj, 769, 88,
  \dodoi{10.1088/0004-637X/769/2/88}

\bibitem[{{Nordquist} {et~al.}(1999){Nordquist}, {Klinger}, {Laguna}, \&
  {Charlton}}]{Nordquist1999}
{Nordquist}, H.~K., {Klinger}, R.~J., {Laguna}, P., \& {Charlton}, J.~C. 1999,
  \mnras, 304, 288, \dodoi{10.1046/j.1365-8711.1999.02293.x}

\bibitem[{{Oliveira} {et~al.}(2020){Oliveira}, {Souza}, {Kerber}, {Barbuy},
  {Ortolani}, {Piotto}, {Nardiello}, {P{\'e}rez-Villegas}, {Maia}, {Bica},
  {Cassisi}, {D'Antona}, {Lagioia}, {Libralato}, {Milone}, {Anderson},
  {Aparicio}, {Bedin}, {Brown}, {King}, {Marino}, {Pietrinferni}, {Renzini},
  {Sarajedini}, {van der Marel}, \& {Vesperini}}]{Oliveira2020}
{Oliveira}, R.~A.~P., {Souza}, S.~O., {Kerber}, L.~O., {et~al.} 2020, \apj,
  891, 37, \dodoi{10.3847/1538-4357/ab6f76}

\bibitem[{{Origlia} {et~al.}(2013){Origlia}, {Massari}, {Rich}, {Mucciarelli},
  {Ferraro}, {Dalessandro}, \& {Lanzoni}}]{Origlia2013}
{Origlia}, L., {Massari}, D., {Rich}, R.~M., {et~al.} 2013, \apjl, 779, L5,
  \dodoi{10.1088/2041-8205/779/1/L5}

\bibitem[{{Origlia} {et~al.}(2011){Origlia}, {Rich}, {Ferraro}, {Lanzoni},
  {Bellazzini}, {Dalessandro}, {Mucciarelli}, {Valenti}, \&
  {Beccari}}]{Origlia2011}
{Origlia}, L., {Rich}, R.~M., {Ferraro}, F.~R., {et~al.} 2011, \apjl, 726, L20,
  \dodoi{10.1088/2041-8205/726/2/L20}

\bibitem[{{Ortolani} {et~al.}(2019){Ortolani}, {Held}, {Nardiello}, {Souza},
  {Barbuy}, {P{\'e}rez-Villegas}, {Cassisi}, {Bica}, {Momany}, \&
  {Saviane}}]{Ortolani2019}
{Ortolani}, S., {Held}, E.~V., {Nardiello}, D., {et~al.} 2019, \aap, 627, A145,
  \dodoi{10.1051/0004-6361/201935726}

\bibitem[{{Pallanca} {et~al.}(2019){Pallanca}, {Ferraro}, {Lanzoni},
  {Saracino}, {Raso}, \& {Focardi}}]{Pallanca2019}
{Pallanca}, C., {Ferraro}, F.~R., {Lanzoni}, B., {et~al.} 2019, \apj, 882, 159,
  \dodoi{10.3847/1538-4357/ab35db}

\bibitem[{{Pallanca} {et~al.}(2021{\natexlab{a}}){Pallanca}, {Lanzoni},
  {Ferraro}, {Casagrande}, {Saracino}, {Purohith Bhaskar Bhat}, {Leanza},
  {Dalessandro}, \& {Vesperini}}]{Pallanca2021b}
{Pallanca}, C., {Lanzoni}, B., {Ferraro}, F.~R., {et~al.} 2021{\natexlab{a}},
  \apj, 913, 137, \dodoi{10.3847/1538-4357/abf938}

\bibitem[{{Pallanca} {et~al.}(2021{\natexlab{b}}){Pallanca}, {Ferraro},
  {Lanzoni}, {Crociati}, {Saracino}, {Dalessandro}, {Origlia}, {Rich},
  {Valenti}, {Geisler}, {Mauro}, {Villanova}, {Moni Bidin}, \&
  {Beccari}}]{Pallanca2021a}
{Pallanca}, C., {Ferraro}, F.~R., {Lanzoni}, B., {et~al.} 2021{\natexlab{b}},
  \apj, 917, 92, \dodoi{10.3847/1538-4357/ac0889}

\bibitem[{{Pietrinferni} {et~al.}(2021){Pietrinferni}, {Hidalgo}, {Cassisi},
  {Salaris}, {Savino}, {Mucciarelli}, {Verma}, {Silva Aguirre}, {Aparicio}, \&
  {Ferguson}}]{Pietrinferni2021}
{Pietrinferni}, A., {Hidalgo}, S., {Cassisi}, S., {et~al.} 2021, \apj, 908,
  102, \dodoi{10.3847/1538-4357/abd4d5}

\bibitem[{{Raso} {et~al.}(2020){Raso}, {Libralato}, {Bellini}, {Ferraro},
  {Lanzoni}, {Cadelano}, {Pallanca}, {Dalessandro}, {Piotto}, {Anderson}, \&
  {Sohn}}]{Raso2020}
{Raso}, S., {Libralato}, M., {Bellini}, A., {et~al.} 2020, \apj, 895, 15,
  \dodoi{10.3847/1538-4357/ab8ae7}

\bibitem[{{Sandage}(1953)}]{Sandage1953}
{Sandage}, A.~R. 1953, \aj, 58, 61, \dodoi{10.1086/106822}

\bibitem[{{Sandell} {et~al.}(1987){Sandell}, {Stevens}, \&
  {Heiles}}]{Sandell1987}
{Sandell}, G., {Stevens}, M.~A., \& {Heiles}, C. 1987, \aap, 179, 255

\bibitem[{{Santos} \& {Piatti}(2004)}]{Santos2004}
{Santos}, J.~F.~C., J., \& {Piatti}, A.~E. 2004, \aap, 428, 79,
  \dodoi{10.1051/0004-6361:20041560}

\bibitem[{{Saracino} {et~al.}(2016){Saracino}, {Dalessandro}, {Ferraro},
  {Geisler}, {Mauro}, {Lanzoni}, {Origlia}, {Miocchi}, {Cohen}, {Villanova}, \&
  {Moni Bidin}}]{Saracino2016}
{Saracino}, S., {Dalessandro}, E., {Ferraro}, F.~R., {et~al.} 2016, \apj, 832,
  48, \dodoi{10.3847/0004-637X/832/1/48}

\bibitem[{{Saracino} {et~al.}(2019){Saracino}, {Dalessandro}, {Ferraro},
  {Lanzoni}, {Geisler}, {Cohen}, {Bellini}, {Vesperini}, {Salaris}, {Cassisi},
  {Pietrinferni}, {Origlia}, {Mauro}, {Villanova}, \& {Moni
  Bidin}}]{Saracino2019}
---. 2019, \apj, 874, 86, \dodoi{10.3847/1538-4357/ab07c4}

\bibitem[{{Schlafly} \& {Finkbeiner}(2011)}]{schlafly11}
{Schlafly}, E.~F., \& {Finkbeiner}, D.~P. 2011, \apj, 737, 103,
  \dodoi{10.1088/0004-637X/737/2/103}

\bibitem[{{Stasi{\'n}ska} {et~al.}(1992){Stasi{\'n}ska}, {Tylenda}, {Acker}, \&
  {Stenholm}}]{Stasinska1992}
{Stasi{\'n}ska}, G., {Tylenda}, R., {Acker}, A., \& {Stenholm}, B. 1992, \aap,
  266, 486

\bibitem[{{Stetson}(1987)}]{Stetson1987}
{Stetson}, P.~B. 1987, \pasp, 99, 191, \dodoi{10.1086/131977}

\bibitem[{{Stetson}(1994)}]{Stetson1994}
---. 1994, \pasp, 106, 250, \dodoi{10.1086/133378}

\bibitem[{{Thompson} {et~al.}(2020){Thompson}, {Udalski}, {Dotter}, {Rozyczka},
  {Schwarzenberg-Czerny}, {Pych}, {Beletsky}, {Burley}, {Marshall},
  {McWilliam}, {Morrell}, {Osip}, {Monson}, {Persson}, {Szyma{\'n}ski},
  {Soszy{\'n}ski}, {Poleski}, {Ulaczyk}, {Wyrzykowski}, {Koz{\l}owski},
  {Mr{\'o}z}, {Pietrukowicz}, \& {Skowron}}]{Thompson2020}
{Thompson}, I.~B., {Udalski}, A., {Dotter}, A., {et~al.} 2020, \mnras, 492,
  4254, \dodoi{10.1093/mnras/staa032}

\bibitem[{{Udalski}(2003)}]{Udalski2003}
{Udalski}, A. 2003, \apj, 590, 284, \dodoi{10.1086/374861}

\bibitem[{{Valenti} {et~al.}(2007){Valenti}, {Ferraro}, \&
  {Origlia}}]{Valenti2007}
{Valenti}, E., {Ferraro}, F.~R., \& {Origlia}, L. 2007, \aj, 133, 1287,
  \dodoi{10.1086/511271}

\bibitem[{{Valenti} {et~al.}(2010){Valenti}, {Ferraro}, \&
  {Origlia}}]{Valenti2010}
---. 2010, \mnras, 402, 1729, \dodoi{10.1111/j.1365-2966.2009.15991.x}

\bibitem[{{Valenti} {et~al.}(2013){Valenti}, {Zoccali}, {Renzini}, {Brown},
  {Gonzalez}, {Minniti}, {Debattista}, \& {Mayer}}]{Valenti2013}
{Valenti}, E., {Zoccali}, M., {Renzini}, A., {et~al.} 2013, \aap, 559, A98,
  \dodoi{10.1051/0004-6361/201321962}

\bibitem[{{Zhang} {et~al.}(2010){Zhang}, {Zhang}, {Han}, \& {Liu}}]{Zhang2010}
{Zhang}, Y., {Zhang}, F., {Han}, Z., \& {Liu}, J. 2010, \apss, 329, 255,
  \dodoi{10.1007/s10509-010-0309-y}

\bibitem[{{Zoccali} {et~al.}(2001){Zoccali}, {Renzini}, {Ortolani},
  {Bragaglia}, {Bohlin}, {Carretta}, {Ferraro}, {Gilmozzi}, {Holberg},
  {Marconi}, {Rich}, \& {Wesemael}}]{Zoccali2001}
{Zoccali}, M., {Renzini}, A., {Ortolani}, S., {et~al.} 2001, \apj, 553, 733,
  \dodoi{10.1086/320980}

\bibitem[{{Zoccali} {et~al.}(2003){Zoccali}, {Renzini}, {Ortolani}, {Greggio},
  {Saviane}, {Cassisi}, {Rejkuba}, {Barbuy}, {Rich}, \& {Bica}}]{Zoccali2003}
---. 2003, \aap, 399, 931, \dodoi{10.1051/0004-6361:20021604}

\end{thebibliography}
\bibliographystyle{aasjournal}



\end{document}